\documentclass[aps,prl,twocolumn,superscriptaddress,10pt,longbibliography]{revtex4-2}

\usepackage{color}
\usepackage{amsmath}
\usepackage{amssymb}
\usepackage{bm}
\usepackage{dsfont}
\usepackage[normalem]{ulem}
\usepackage{float}

\usepackage[T1]{fontenc}

\usepackage{circledsteps}
\usepackage{hyperref}

\usepackage{graphicx}
\graphicspath{{figs/}}

\newcommand{\beq}{\begin{eqnarray}}
\newcommand{\eeq}{\end{eqnarray}} 

\newcommand{\hide}[1]{}

\newcommand{\sect}[1]{{\it #1.-- }}

\newcommand{\Cn}[1]{\begin{center} #1 \end{center}}

\hypersetup{
  colorlinks=true,
  allcolors = blue,
  pdftitle={}
}

\makeatletter
\def\maketitle{
\@author@finish
\title@column\titleblock@produce
\suppressfloats[t]}
\makeatother

\begin{document}
	
\title{Stabilizing open photon condensates by ghost-attractor dynamics}

    \author{Aya Abouelela$^a$}
        \email{ayaaabou@uni-bonn.de}
        \affiliation{Physikalisches Institut, Universit\"at Bonn, Nussallee 12, 53115, Bonn, Germany}

    \author{Michael Turaev$^a$}
        \email{mturaev@uni-bonn.de}
        \affiliation{Physikalisches Institut, Universit\"at Bonn, Nussallee 12, 53115, Bonn, Germany}
       
	\author{Roman Kramer}
        \affiliation{Physikalisches Institut, Universit\"at Bonn, Nussallee 12, 53115, Bonn, Germany}

    \author{\hbox{Moritz Janning}}
        \affiliation{Physikalisches Institut, Universit\"at Bonn, Nussallee 12, 53115, Bonn, Germany}

    \author{Michael Kajan}
        \affiliation{Physikalisches Institut, Universit\"at Bonn, Nussallee 12, 53115, Bonn, Germany}

    \author{Sayak Ray}
        \email{sayak@uni-bonn.de}
        \affiliation{Physikalisches Institut, Universit\"at Bonn, Nussallee 12, 53115, Bonn, Germany}

    \author{Johann Kroha}
	\email{jkroha@uni-bonn.de}
        \affiliation{Physikalisches Institut, Universit\"at Bonn, Nussallee 12, 53115, Bonn, Germany}
        \affiliation{\hbox{School of Physics and Astronomy, University of St. Andrews, North Haugh, St. Andrews, KY16 9SS, United Kingdom}}

        \begin{abstract}
        \vspace*{-0.4cm}
        \centerline{\hbox{$^a$These authors contributed equally to this work.}}
        \vspace*{0.3cm}       
            We study the temporal, driven-dissipative dynamics of open photon Bose-Einstein condensates (BEC) in a dye-filled microcavity, taking the condensate amplitude and the noncondensed fluctuations into account on the same footing by means of a cumulant expansion within the Lindblad formalism. The fluctuations fundamentally alter the dynamics in that the BEC always dephases to zero for a sufficiently long time. However, a \textit{ghost attractor}, although outside of the physically accessible configuration space, attracts the dynamics and leads to a plateau-like stabilization of the BEC for an exponentially long time, consistent with experiments. We also show that the photon BEC and the lasing state are separated by a true phase transition, since they are characterized by different fixed points. The ghost-attractor nonequilibrium stabilization mechanism is an alternative to prethermalization and may possibly be realized on other dynamical platforms as well.   
        \end{abstract}
	
	
\maketitle
	
Quantum gases of photons have proven to be a versatile platform for investigating various quantum effects in many-body systems \cite{Ciuti13}, including Bose-Einstein condensation \cite{Weitz10, Nyman15, Oosten18, Wouter22, Nyman-rev18} and its dimensional crossover \cite{Kiran24}, quantum coherence \cite{Nyman18, Schmitt16}, thermodynamic properties of driven-dissipative systems \cite{Schmitt22, Schmitt23} and even a nonHermitian phase transition within the Bose-Einstein condensed phase \cite{Ozturk21}.  Bose-Einstein condensates (BEC) of photons have been realized as photon gases in dye-filled, optical cavities, where the discreteness of the longitudinal cavity spectrum induces a nonzero effective mass to the cavity photons \cite{Weitz10, Nyman15, Oosten18}, and most recently in vertical-cavity surface-emitting laser (VCSEL) systems \cite{Pieczarka24}. Due to repeated photon emission and reabsorption processes by the dye molecules and energy exchange with their vibrational excitations, the photon gas reaches a near-thermal state, which can ultimately undergo a Bose-Einstein condensation transition. 

Being a driven-dissipative system, the photon BEC is naturally prone to decay due to various dephasing processes, like grand-canonical exchange of excitations with the dye-molecule bath, cavity loss, and normal (noncondensed) photon-density fluctuations \cite{Schmitt14, Nyman16, Nyman24, TimBode19}.  
However, due to the complexity of its spatio-temporal dynamics, the photon gas is often treated at the mean-field level \cite{Wouter20,Gladlin20}, neglecting the normal photon fluctuations. 
Conversely, dynamics of the photon gas has been studied in Refs.~\cite{Keeling13,Keeling15,Pelster,Nyman19,TimBode24} by considering the total photon density without distinguishing between the $U(1)$-symmetry breaking condensate and the normal photon density. Investigations involving both the condensate field and its fluctuations, which are crucial for understanding the conditions of stability and dephasing of the open photon condensate, have remained scarce \cite{Keeling-symmetry-breaking}. 

In this Letter, we present a systematic study of the coupled, temporal dynamics of continuously driven, dissipative photon BECs, their noncondensed fluctuations, and the dye-molecule excitations, based on the Lindblad formalism and treating all dynamical fields on the same footing by means of a second-order cumulant expansion. The interplay of symmetry-breaking and noncondensed fields change the dynamics fundamentally compared to Gross-Pitaevskii-like dynamics \cite{Wouter20,Gladlin20}.  In the weakly driven regime, the photon BEC always decays to zero in the long-time limit. However, at finite times the dynamics are governed by a previously undiscovered \textit{ghost attractor}, a concept known from nonlinear systems \cite{Strogatz-book,Strogatz89,Koch24} and realized here as a fixed point (FP) in configuration space which exists outside of the physically accessible realm with an unphysical condensate fraction, $\nu > 1$.
The system dynamics evolve towards this FP, but stall due to its inaccessibility. 
This leads to a constant, plateau-like photon BEC amplitude, before it is repelled due to a positive Lyapunov exponent, and then exponentially decays to zero. We estimate the decay time by linear stability analysis as well as numerical time evolution. Generically, the plateau lifetime is macroscopically large, consistent with the experimental observation of a stationary photon BEC over the entire observation time \cite{Weitz10,Schmitt16}. Entering the strongly pumped regime, the ghost FP shifts away from the physical regime even more, destabilizing the photon BEC, and a new, physical FP with $\nu<1$ appears, which is a state with population inversion of the molecule excitations and an infinitely long-lived photon condensate amplitude. This change of stable FPs thus marks a nonequilibrium quantum phase transition from a photon BEC to the lasing regime.  

\begin{figure}[t]
\includegraphics[width=\columnwidth]{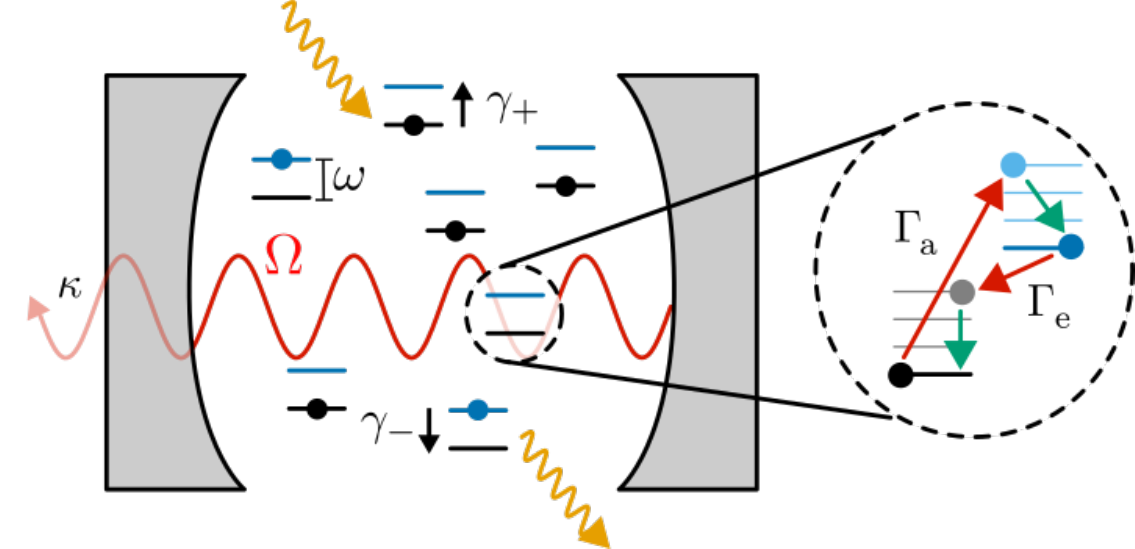}
\caption{{\it Sketch of the setup.} The microcavity of single-mode frequency $\Omega$ is filled with dye molecules of electronic transition frequency $\omega$. The inset shows fast relaxation of the vibrational modes of molecules (green arrows). The dissipative processes include the absorption (emission) rates $\Gamma_{\rm a}$ $(\Gamma_{\rm e}$), the cavity-photon loss $\kappa$, the external pumping and nonradiative decay of the excited molecules $\gamma_{\pm}$, respectively.}
\label{fig:model-schematic}
\end{figure}

\sect{Model and formalism}
The coupled system of cavity photons and dye molecules immersed in a solvent is described in a standard way by a Jaynes-Cummings model with additional molecular vibrational modes (phonons) as sketched in Fig.~\ref{fig:model-schematic} \cite{Keeling13,Keeling15,Pelster,TimBode19}. 
Photon emission and absorption induce dipole transitions between the molecular electronic ground and excited states, respectively, which, in turn, couple to the phonon excitations via the Franck-Condon effect \cite{Condon1928}. 
A polaron transformation diagonalizes the molecular part of the Hamiltonian and induces a renormalized, nonlinear Jaynes-Cummings coupling. Due to fast solvent collisions, the phonon subsystem is, to very good approximation, in thermal equilibrium. Treating, thus, the phonons as a thermal bath, one arrives at the Lindblad master equation for the density matrix of the coupled system of cavity photons and electronic excitations \cite{Petruccione_book,Keeling13,TimBode19},
\begin{eqnarray}
    \partial_t \hat{\rho} &=& -i \left[\hat{H},\hat{\rho}\right] + \sum_{j=1}^M \left(\Gamma_{\rm a} \mathcal{L}[\hat{a} \hat{\sigma}_j^+] + \Gamma_{\rm e} \mathcal{L}[\hat{a}^\dagger \hat{\sigma}_j^-]\right) \nonumber \\
    &+& \sum_{j=1}^M \left(\gamma_{+} \mathcal{L}[\hat{\sigma}_j^+] + \gamma_{-} \mathcal{L}[\hat{\sigma}_j^-] \right) + \kappa \mathcal{L}[\hat{a}].
\label{eq:LME}
\end{eqnarray}
Here, the photons in a single cavity mode are represented by the bosonic annihilation (creation) operators $\hat{a}$ ($\hat{a}^{\dagger}$) and the two-level system of electronic ground and excited states of molecule number $j$ by the Pauli matrix $\hat\sigma_{j}^z$ and the raising/lowering operators $\hat\sigma_j^{\pm}=(\hat{\sigma}_{j}^x \pm i \hat{\sigma}_{j}^y)/2$.
The Lindblad superoperator associated with an operator $\hat{\mathcal{O}}$ and acting on the density matrix $\hat\rho$ reads, $\mathcal{L}[\hat{\mathcal{O}}]\hat\rho= \hat{\mathcal{O}} \hat\rho \hat{\mathcal{O}}^\dagger - \left(\hat{\mathcal{O}}^\dagger \hat{\mathcal{O}} \hat\rho + \hat\rho \hat{\mathcal{O}}^\dagger \hat{\mathcal{O}}\right)/2$.  
The external laser pump and the radiation-less decay rate are represented by $\gamma_+$ and $\gamma_-$, respectively ($\gamma_- \ll \gamma_+$), $\kappa$ parametrizes the cavity-photon loss, and $M$ is the total number of dye molecules. $\Gamma_a$ and $\Gamma_e$ are the renormalized, phonon-assisted Einstein absorption and emission coefficients, respectively. Since the phonon bath is thermal, they obey the Kennard-Stepanov relation, $\Gamma_{\rm a}=\Gamma_{\rm e}e^{\beta \delta}$ \cite{Keeling15}, where $\beta$ is the inverse temperature of the bath. 

The coherent time-evolution in Eq.~\eqref{eq:LME} is governed by the Hamiltonian $\hat{H}$ in the rotating frame, 
\begin{equation}
\hat{H} = \delta \hat{a}^\dagger \hat{a} + g_\beta  \sum_{j=1}^M \left( \hat{a}^\dagger \hat{\sigma}_j^- + \hat{a} \hat{\sigma}_j^+ \right),
\label{JCH}
\end{equation}
where $\delta=\Omega-\omega$ is the detuning of the cavity frequency $\Omega$ from the transition frequency $\omega$ of the electronic excitations, and $g_{\beta}$ is the Jaynes-Cummings coupling renormalized by the polaron transformation \cite{Pelster}.
Note that in experiments, the absorption or emission coefficients $\Gamma_a$, $\Gamma_e$ can be sensitively tuned by the detuning $\delta$, but the detuning term in $\hat{H}$ only contributes an irrelevant, global phase rotation to the symmetry-breaking quantities.

\sect{Equations of motion}
Nonlinear rate equations for the single-time expectation value of any operator $\mathcal{O}$ are derived from the master equation \eqref{eq:LME} as $\partial_t \langle \hat{\mathcal{O}} \rangle = \mathrm{Tr}[\hat{\mathcal{O}} \partial_t \hat{\rho}]$. 
To describe both the condensate and the fluctuation dynamics, we consider all expectation values up to second order in the photon and molecule operators, namely
the U(1)-symmetry-breaking amplitudes $\psi(t):=\langle \hat a\rangle$, $\chi(t):=\langle \hat\sigma_{j}^{-}\rangle$, the fraction of excited molecules $m_e(t):=(1+\langle\hat\sigma_j^z\rangle)/2$, and the cavity-photon number $n(t):=\langle \hat a^{\dagger}\hat a\rangle$, as well as $\langle \hat a \hat a\rangle$, $\langle \hat a^{\dagger}\hat \sigma_{j}^\alpha\rangle$, $\langle \hat \sigma_{j}^\alpha \hat \sigma_k^{\beta}\rangle$, with $\alpha,\beta=+,-,z$. 

(i) We systematically reduce higher-order correlators generated by the master equation to products of these expectation values by means of a cumulant expansion. 
(ii) Intermolecular correlations are efficiently destroyed by fast solvent collisions in our dilute dye solution. Therefore, nonlocal expectation values factorize as, e.g., $\langle\hat\sigma_j^+ \hat\sigma_k^- \rangle = 
\delta_{jk} (1+\langle \hat \sigma_j^z\rangle )/2  + (1-\delta_{jk}) \langle \hat\sigma_j^+\rangle \langle \hat\sigma_k^-\rangle$. Since the photonic wavelength is much larger than the molecule spacing, we can further assume spatial homogeneity of the molecule degrees of freedom, so that we will write, $\langle\hat\sigma_j^\alpha\rangle=\langle\hat\sigma_k^\alpha\rangle=:\langle\hat\sigma^\alpha\rangle$. (iii) Polaritonic correlators between photonic and molecular degrees of freedom factorize, e.g., $\langle \hat a^{\dagger}\hat\sigma^{-}\rangle = \langle \hat a^{\dagger}\rangle \langle\hat\sigma^-\rangle$, since we work in a regime of weak photon-molecule coupling, where polaritonic states \cite{Keeling22, Keeling-polariton20} are not formed, i.e., the photons are independent particles. We verified this by deriving and explicitly solving the equation of motion for $\langle\hat a^{\dagger}\hat\sigma^-\rangle$. Furthermore, the equation of motion for $\langle\hat{a}\hat{a}\rangle$ decouples from those of the other quantities listed above. The dynamical equations then reduce to the following set of four coupled, nonlinear rate equations,   
\begin{subequations}
\begin{eqnarray}
\partial_t{\psi} &=& -ig_{\beta} M \chi - \left[\kappa - MG_{\psi}(m_e)\right]\psi/2 
\label{eq:psit}
\\
\partial_t{\chi} &=& i g_{\beta} \psi (2m_e-1) - G_{\chi}(n)\chi/2
\label{eq:chit}
\\
\partial_t{m}_e &=& - 2 g_{\beta} {\rm Im}[\psi^* \chi] + \gamma_{+}(1-m_e) - \gamma_{-} m_e \nonumber \\
&-& R(m_e,n)
\label{eq:met}
\\
\partial_t{n} &=& 2 g_{\beta} M {\rm Im}[\psi^* \chi] - \kappa n + MR(m_e,n)\ ,
\label{eq:nt}
\end{eqnarray}
\label{eq:rate_eqn}
\end{subequations}
where the molecule-induced dissipation rates of the amplitudes $\psi$, $\chi$ read, $G_{\psi}(m_e)=\Gamma_{\rm e}m_e - \Gamma_{\rm a}(1-m_e)$ and $G_{\chi}(n)=(\gamma_{+}+\gamma_{-}) + \Gamma_{\rm a} n + \Gamma_{\rm e} (n+1)$, respectively. The molecule-induced photon gain/loss is given by $R(m_e,n)=\Gamma_{\rm e} (n+1)m_e - \Gamma_{\rm a} n (1-m_e)$. Rewriting $R(m_e,n)=nG_{\psi}(m_e)+\Gamma_{\rm e}m_e$, one may identify the spontaneous-emission term $\Gamma_{\rm e}m_e$, which is not present in the condensate equation of motion \eqref{eq:psit}.
With the appearance of the U(1)-symmetry-breaking condensate field $\psi$, the molecular emission amplitude $\chi$ necessarily appears.
It may be nonzero because of the nonorthogonality of the effective molecule ground and excited states after polaron transformation.       
Due to the positivity of the molecule (pseudospin) density matrix, $\chi$ is bounded by $|\chi| \leq \chi_{\rm max}=\sqrt{m_e(1-m_e)}$ \cite{Laussy16}. 

\sect{Fixed points and their stability} 
We begin by analyzing the steady-state solutions $\overline{X}=(\overline{\psi},\overline{\chi},\overline{m}_e,\overline{n})$ of the system \eqref{eq:rate_eqn} obtained by setting the time derivatives equal to zero.
Expressing $\overline{\chi}$ in terms of $\overline{\psi}$ using Eq.~\eqref{eq:psit}, inserting it into Eq.~\eqref{eq:nt}, and using the expression for $R(\overline{m}_e,\overline{n})$, the condensate fraction reads,
\begin{equation}
\overline{\nu} = \frac{|\overline{\psi}|^2}{\overline{n}} = 
1-\frac{M\Gamma_e \overline{m}_e}{\overline{n}[\kappa-MG_{\psi}(\overline{m}_e)]}\ , 
\label{eq:nu}
\end{equation}
and the general relation 
\begin{equation}
\overline{m}_e = \left( \gamma_+ - \frac{\kappa \overline{n}}{M} \right)/(\gamma_+ + \gamma_-)
\label{eq:m_e-n}
\end{equation}
holds.  As seen from Eqs.~\eqref{eq:psit} and \eqref{eq:chit}, there exists a 
steady-state solution $\overline{X}^0=(0,0,\overline{n}^0,\overline{m}_e^0)$ with vanishing condensate and molecule emission amplitudes. Its photon number $\overline{n}^0$ and excitation fraction $\overline{m}_e^0$ can be found from Eqs.~\eqref{eq:met} and \eqref{eq:nt}. 
In addition, there exists a solution with nonvanishing $\overline{\psi}$. Since Eqs.~\eqref{eq:psit} and \eqref{eq:chit} are homogeneous and linear in $\overline{\psi}$ and $\overline{\chi}$, this requires that the determinant of these equations vanishes, 
\begin{equation}
    4 g_{\beta}^2 M (2\overline{m}_e-1) - G_{\chi}(\overline{n}) \left[\kappa-MG_{\psi}(\overline{m}_e)\right] = 0.
\label{eq:determinant} 
\end{equation}
We first consider the experimentally relevant \cite{Weitz10}, weakly pumped regime with no population inversion of molecule excitations, $2\overline{m}_e-1< 0$. 
The fixed-point values $\overline{n}=\overline{n}^{\rm G}$ and $\overline{m}_e=\overline{m}_e^{\rm G}$ are obtained from Eqs.~\eqref{eq:m_e-n} and \eqref{eq:determinant}, and Eq.~\eqref{eq:chit} with \eqref{eq:nt} then determine the amplitudes $\overline{\psi}=\overline{\psi}^{\rm G}$ and $\overline{\chi}=\overline{\chi}^{\rm G}$ uniquely up to an irrelevant, global phase factor. 
The value $\overline{m}_e^{\rm G}$ satisfies $\left[\kappa - MG_{\psi}(\overline{m}_e^{\rm G})\right]<0$, implying $\overline{\nu}=\overline{\nu}^{\rm G} >1$ in Eq.~\eqref{eq:nu}, an unphysical condensate fraction. 
We, therefore, dub this solution a \textit{ghost fixed point}, $\overline{X}^{\rm G} = (\overline{\psi}^{\rm G},\overline{\chi}^{\rm G},\overline{m}_e^{\rm G},\overline{n}^{\rm G})$. Although it cannot be physically realized, it does crucially influence the time-dependent dynamics, which we study next. 

\begin{figure}
\includegraphics[width=0.9\columnwidth]{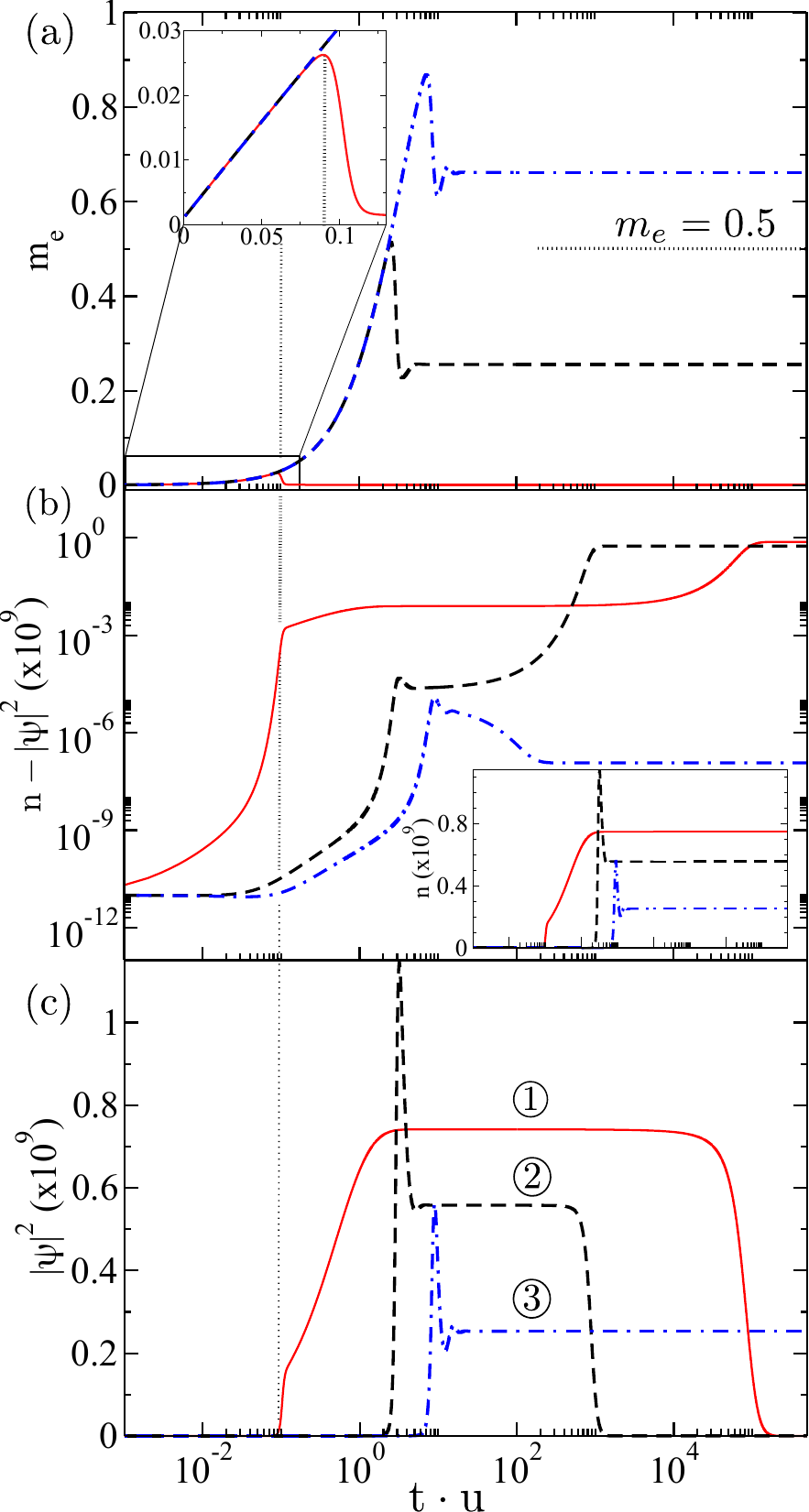}
\caption{{\it Time evolution.} Relaxation dynamics of the fraction of excited molecule $m_{e}$, the photon fluctuation $n-|\psi|^2$, and the photon condensate density $|\psi|^2$ are shown in (a-c), respectively, and the photon number $n$ as inset in (b), for different ratios of absorption and emission, $\log_{10}\left(\Gamma_{\rm a}/\Gamma_{\rm e}\right)=$ \Circled{1} $-3.25$, \Circled{2} $-0.65$ and \Circled{3} $-0.04$.
The onset of condensation in \Circled{1} is accompanied by molecule de-excitation in $m_e$, see also inset of (a).
The other parameters values used here are, $\kappa/u=2$, $\gamma_{-}/u=5\times 10^{-5}$, $\gamma_{+}/u= 0.3$, $\Gamma_{\rm a}/u=10^{-9}$ and $M=5\times 10^9$\textcolor{red}, consistent with experimental parameters~\cite{schmitt_absorption_2024, Pelster}. Here and in the other figures of this paper we scale energy (time) by a unit $u$ ($1/u$), defined as $u=10^{-4}Mg_{\beta}$.}
\label{fig:time_evol}
\end{figure}

The behavior near the FPs can be analyzed by linear stability analysis, which involves computing the stability matrix $\mathcal{M}$ and its eigenvalues $\lambda$ by parameterizing the deviations from $\overline{X}$ as $X(t)=\overline{X}+\Delta X\, e^{\lambda t}$, see \cite{Supplement}.  
The detailed, analytical fixed-point solutions sketched above show that the noncondensed FP $\overline{X}^0$ is stable, that is, ${\rm Re}(\lambda)<0$ for all eigenvalues, while the ghost FP $\overline{X}^{\rm G}$ is metastable in that for one of the eigenvalues ${\rm Re}(\lambda)>0$ and ${\rm Re}(\lambda)<0$ otherwise, see \cite{Supplement}. 

\sect{Temporal dynamics and stabilization by ghost} 
For a typical set of parameters, the resulting dynamics of the excited molecule fraction $m_e$, the photon fluctuation number $n-|\Psi|^2$, and the condensate photon number $|\psi|^2$ are plotted in Fig.~\hyperref[fig:time_evol]{\ref*{fig:time_evol}(a-c)}, respectively, for different ratios of absorption and emission coefficients, $\Gamma_{\rm a}/\Gamma_{\rm e}$. At time $t=0$, we prepared the system with an almost vanishing photon density, $n(0)/M\ll 1$, and with $m_e(0)=0.1\%$ excited molecules. For the emission amplitude, we assumed the initial value $\chi(0)=\chi_{\rm max}$. The dependence of the dynamics on $\chi (0)$ is discussed in the End Matter.
\begin{figure}[H]
\includegraphics[width=\columnwidth]{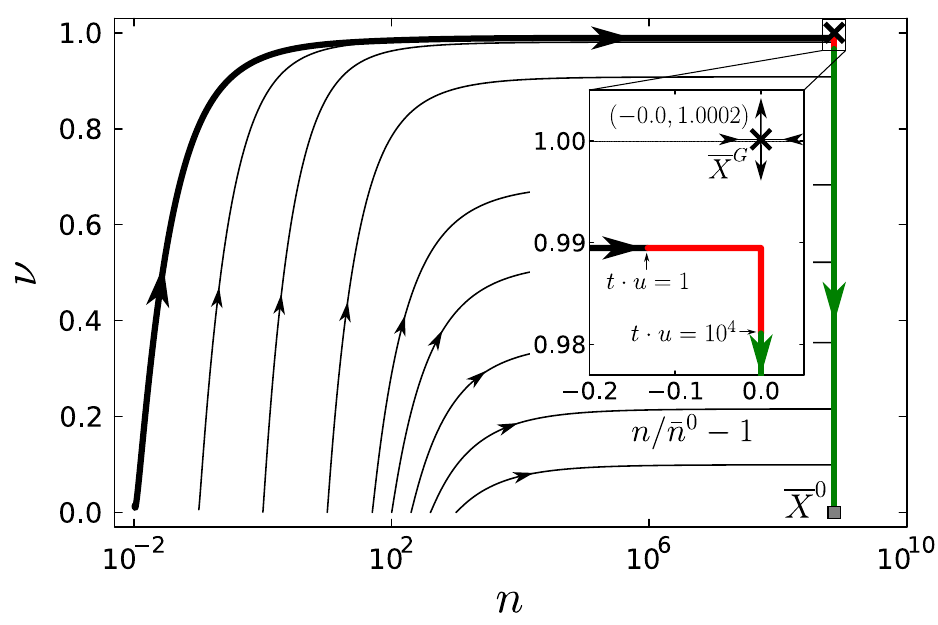}
\caption{{\it Flow diagram.} Projections of the phase-space trajectories are plotted in the $n-\nu$ plane for several initial values $n(0)$. The thick line represents the time-evolution \Circled{1} in Fig.~\ref{fig:time_evol}, and the time-frames are indicated by different colors (see the inset). The FPs $\overline{X}^{\rm G}$ and $\overline{X}_{\rm FP}^{0}$ are marked by a cross and a filled square, respectively, for reference. 
The dynamics in the vicinity of $\overline{X}^{\rm G}$ is shown in the inset, with the eigenvectors corresponding to the smallest negative (attractive) and largest positive (repulsive) eigenvalues of $\mathcal{M}$. 
}
\label{fig:flow-diagram}
\end{figure}

There are several time scales involved in the dynamics. The initial time period is dominated by an exponential growth of the noncondensed cavity-photon number $n(t)-|\psi(t)|^2$, whose end is marked by a sudden population of the photon BEC. Initially, $m_e(t)$ increases, primarily governed by $\partial_t{m_e} \approx \gamma_+(1-m_e)$ in Eq.~\eqref{eq:met}, until the condensate population sets in. The latter then enters into a metastable plateau region with almost constant $|\psi(t)|^2$, while $m_e(t)$ and $n(t)$ reach constant, asymptotic values (see Figs.~\hyperref[fig:time_evol]{\ref*{fig:time_evol}(a)} and inset of \hyperref[fig:time_evol] {\ref*{fig:time_evol}(b)}), consistent with the previous works \cite{Keeling13, Nyman19}. The plateau duration increases as the cavity detuning $\beta \delta \sim {\log}_{10}(\Gamma_a/\Gamma_e)<0$ is made more negative. For small molecule excitation fraction, $m_e(t)\lesssim 0.1\%$, it reaches values many orders of magnitude larger than a photon emission cycle $1/g_{\beta}$, consistent with experiments \cite{Weitz10,Ozturk21}, before $\psi(t)$ dephases and decays exponentially (cf. Fig.~\hyperref[fig:time_evol]{\ref*{fig:time_evol}(c)}, \Circled{1} and \Circled{2}). The metastable plateau behavior can be understood from the flow diagram in Fig.~\ref{fig:flow-diagram} as due to the ghost FP. For the parameters chosen, $\overline{X}^{\rm G}$ lies at  $\overline{\nu}^{\rm G}\approx 1.0002$, outside of, but close to the physical realm. Due to the attractive components of the stability matrix $\mathcal{M}$ \cite{Supplement}, the system evolves towards $\overline{X}^{\rm G}$, but then essentially stalls because $\overline{X}^{\rm G}$ cannot be reached. The red branch in the inset of Fig.~\ref{fig:flow-diagram} exemplifies that the trajectory spends $t\cdot u \sim 10^4$ in the vicinity of the ghost FP with nearly constant condensate amplitude $\psi(t)$. For larger times, the system is ultimately repelled from $\overline{X}^{\rm G}$ due to a single, positive Lyapunov exponent, ${\rm Re}(\lambda^{\rm G})>0$ \cite{Supplement}, and approaches the stable FP $\overline{X}^0$ with vanishing
\begin{figure}[t]
\includegraphics[width=\columnwidth]{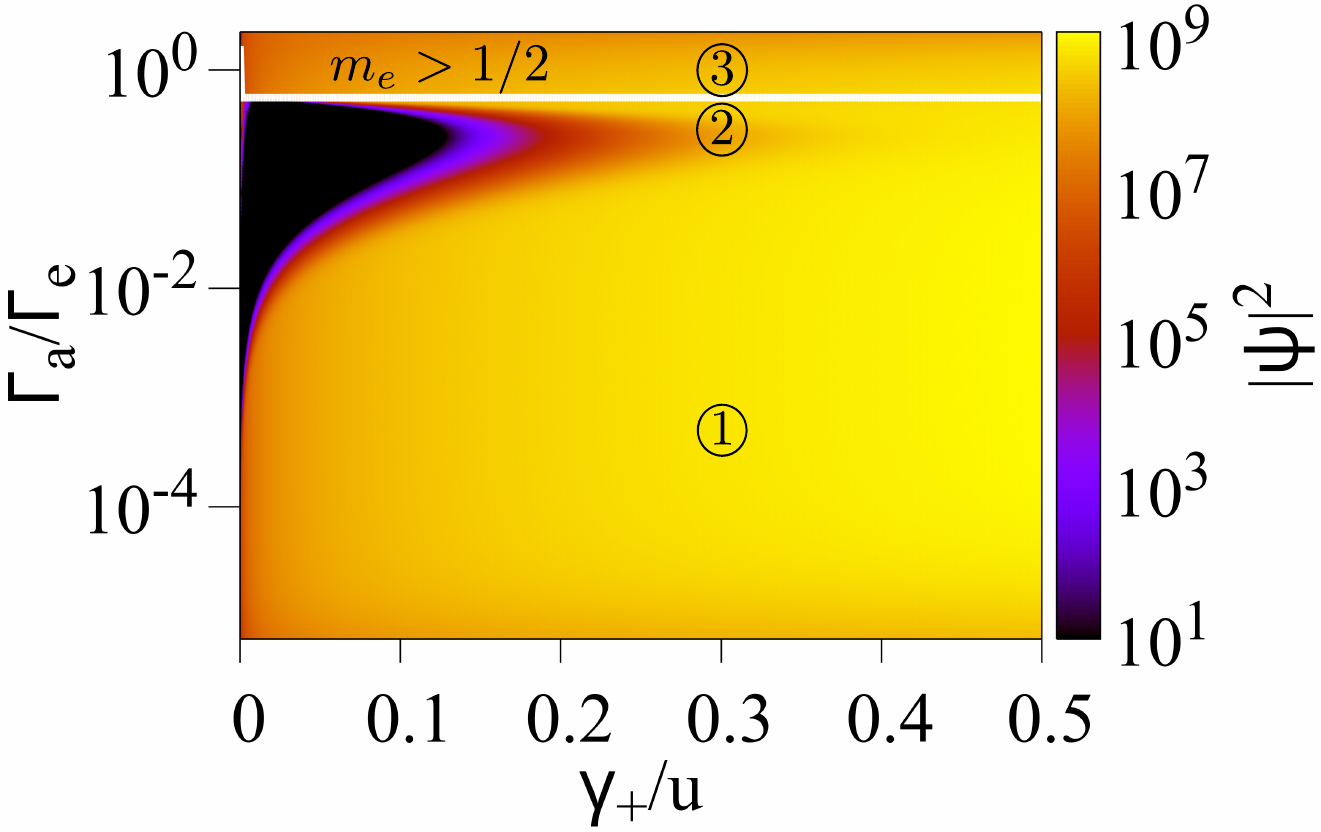}
\caption{{\it Phase diagram.} The condensate photon number $|\psi|^2$ measured at a time $t_0=10^3(1/u)$ (cf. Fig.~\ref{fig:time_evol}) is plotted in a color scale by varying the pump rate $\gamma_{+}/u$ and $\Gamma_{\rm a}/\Gamma_{\rm e}$ for the same parameters as in Fig.~\ref{fig:time_evol}. Representative dynamics of various quantities at \Circled{1} - \Circled{3} are demonstrated in Fig.~\ref{fig:time_evol} and in \cite{Supplement} for lower photon numbers. The lasing region with $m_e>1/2$ is bounded from below by a white line.}
\label{fig:phase_diagram}
\end{figure}
condensate amplitude. The numerically determined plateau time-span agrees well with $1/{\rm Re}(\lambda^{\rm G})$, see End Matter and \cite{Supplement}.

\sect{Phase diagram and transition to laser} 
We refer to the long-lived plateau state with nonzero condensate amplitude, $\psi(t)\neq 0$, but without molecule population inversion, $m_e(t)\ll 1/2$, as an open photon BEC. In our model, we can also tune the system to a population-inverted state by making the detuning $\beta\delta\sim \log_{10}(\Gamma_a/\Gamma_e)$ less negative, which corresponds, by definition, to a lasing state. 
As $\Gamma_a/\Gamma_e$ is increased within the photon-BEC phase, the FP $\overline{X}^0$ with $\overline{\psi}=\overline{\chi}=0$ becomes unstable when population inversion ($2\overline{m}_e-1>0$) is reached, see Fig.~\hyperref[fig:lasing-transition]{\ref*{fig:lasing-transition}(a)}.
At the same time, the ghost FP $\overline{X}^{\rm G}$ shifts further into the unphysical region, $\overline\nu^{\rm G} \to \infty$, as shown in Fig.~\hyperref[fig:lasing-transition]{\ref*{fig:lasing-transition}(b)}.
As a result, the plateau duration decreases and the photon BEC dephases earlier, see \Circled{2} in Fig.~\hyperref[fig:time_evol]{\ref*{fig:time_evol}(c)}.
Instead, when the laser phase boundary is crossed, a new, stable FP $\overline{X}^{\rm L}$ with nonzero $\overline{\psi}=\overline{\psi}^{\rm L}$ and $\overline{\chi}=\overline{\chi}^{\rm L}$ appears, which resides in the physical realm, $\overline{\nu}=\overline{\nu}^{\rm L}<1$, see Fig.~\hyperref[fig:lasing-transition]{\ref*{fig:lasing-transition}(b)}. The dynamics of the corresponding lasing state, i.e., a condensate with population inversion and infinite lifetime, is shown by \Circled{3} in Fig.~\ref{fig:time_evol}. The appearance of two distinct fixed points $\overline{X}^G$, $\overline{X}^L$ characterizing the photon BEC and the lasing state, respectively, shows that the two phases are separated by a true nonequilibrium quantum phase transition. This behavior is shown by the phase diagram of cavity detuning ${\log}_{10}(\Gamma_a/\Gamma_e)$ vs. pump rate $\gamma_+/u$ in Fig.~\ref{fig:phase_diagram}. For moderate pump $\gamma_{+}/u\lesssim 0.1$, the condensate $\psi(t_0)$ calculated at a given time $t_0$ first decreases with increasing ${\log}_{10}(\Gamma_a/\Gamma_e)$, before $\psi(t_0)$ is recovered, as the system enters into the lasing phase marked by the white line. Only for stronger pumping, $|\psi(t_0)|^2$ shows a continuous change between the two phases.     

\sect{Conclusion}
Our results show that, in contrast to pure mean-field dynamics, including noncondensed fluctuations on the same footing drastically changes the dynamics even in the Markovian regime. 
Although the photon BEC always dephases in the infinite-time limit due to its coupling to the noncondensed photon cloud. The condensate dynamics, being attracted to the inaccessible ghost FP, stalls for a long, intermediate time period, longer than the observation time of contemporary experiments \cite{Weitz10, Nyman15, Kiran24, Oosten18, Nyman-rev18, Nyman18, Schmitt16}. 
The ghost-attractor stabilization mechanism for nonequilibrium quantum states is fundamentally distinct from prethermalization, which is, by contrast, based on conservation laws and generalized Gibbs ensembles \cite{Moeckel08,Eckstein11,Essler15,Rigol19,Ray20}. We also identified the relation between the driven-dissipative photon BEC (a condensate without molecule population inversion) and the lasing state (with population inversion) as a true nonequilibrium phase transition characterized by two distinct fixed points of the dynamics. Observable signatures such as different fluctuation spectra, will be a subject of future research. Our findings can be tested experimentally by interference measurements of the condensate amplitude and may be realized in various platforms of light-matter interaction, including polariton condensates \cite{Littlewood19,Keeling22,Keeling-polariton20} or Dicke-like systems \cite{Kollath, Ray24}.

\sect{Acknowledgements}
We thank Eugene Demler, Axel Pelster, Julian Schmitt, and Martin Weitz for useful discussions. This work was funded by the Deutsche Forschungsgemeinschaft (DFG) under Germany's Excellence Strategy-Cluster of Excellence Matter and Light for Quantum Computing, ML4Q (No. 390534769) and through the DFG Collaborative Research Center CRC 185 OSCAR (No. 277625399). S.R. acknowledges a scholarship from the Alexander von Humboldt (AvH) Foundation, Germany.

The data that support the findings of this article are openly available \cite{Zenodo-data_2025}.

\bibliography{refs}

\providecommand{\noopsort}[1]{}\providecommand{\singleletter}[1]{#1}%
\begin{thebibliography}{44}%
\makeatletter
\providecommand \@ifxundefined [1]{%
 \@ifx{#1\undefined}
}%
\providecommand \@ifnum [1]{%
 \ifnum #1\expandafter \@firstoftwo
 \else \expandafter \@secondoftwo
 \fi
}%
\providecommand \@ifx [1]{%
 \ifx #1\expandafter \@firstoftwo
 \else \expandafter \@secondoftwo
 \fi
}%
\providecommand \natexlab [1]{#1}%
\providecommand \enquote  [1]{``#1''}%
\providecommand \bibnamefont  [1]{#1}%
\providecommand \bibfnamefont [1]{#1}%
\providecommand \citenamefont [1]{#1}%
\providecommand \href@noop [0]{\@secondoftwo}%
\providecommand \href [0]{\begingroup \@sanitize@url \@href}%
\providecommand \@href[1]{\@@startlink{#1}\@@href}%
\providecommand \@@href[1]{\endgroup#1\@@endlink}%
\providecommand \@sanitize@url [0]{\catcode `\\12\catcode `\$12\catcode `\&12\catcode `\#12\catcode `\^12\catcode `\_12\catcode `\%12\relax}%
\providecommand \@@startlink[1]{}%
\providecommand \@@endlink[0]{}%
\providecommand \url  [0]{\begingroup\@sanitize@url \@url }%
\providecommand \@url [1]{\endgroup\@href {#1}{\urlprefix }}%
\providecommand \urlprefix  [0]{URL }%
\providecommand \Eprint [0]{\href }%
\providecommand \doibase [0]{https://doi.org/}%
\providecommand \selectlanguage [0]{\@gobble}%
\providecommand \bibinfo  [0]{\@secondoftwo}%
\providecommand \bibfield  [0]{\@secondoftwo}%
\providecommand \translation [1]{[#1]}%
\providecommand \BibitemOpen [0]{}%
\providecommand \bibitemStop [0]{}%
\providecommand \bibitemNoStop [0]{.\EOS\space}%
\providecommand \EOS [0]{\spacefactor3000\relax}%
\providecommand \BibitemShut  [1]{\csname bibitem#1\endcsname}%
\let\auto@bib@innerbib\@empty
\bibitem [{\citenamefont {Carusotto}\ and\ \citenamefont {Ciuti}(2013)}]{Ciuti13}%
  \BibitemOpen
  \bibfield  {author} {\bibinfo {author} {\bibfnamefont {I.}~\bibnamefont {Carusotto}}\ and\ \bibinfo {author} {\bibfnamefont {C.}~\bibnamefont {Ciuti}},\ }\bibfield  {title} {\bibinfo {title} {{Quantum fluids of light}},\ }\href {https://doi.org/10.1103/RevModPhys.85.299} {\bibfield  {journal} {\bibinfo  {journal} {Rev. Mod. Phys.}\ }\textbf {\bibinfo {volume} {85}},\ \bibinfo {pages} {299} (\bibinfo {year} {2013})}\BibitemShut {NoStop}%
\bibitem [{\citenamefont {Klaers}\ \emph {et~al.}(2010)\citenamefont {Klaers}, \citenamefont {Schmitt}, \citenamefont {Vewinger},\ and\ \citenamefont {Weitz}}]{Weitz10}%
  \BibitemOpen
  \bibfield  {author} {\bibinfo {author} {\bibfnamefont {J.}~\bibnamefont {Klaers}}, \bibinfo {author} {\bibfnamefont {J.}~\bibnamefont {Schmitt}}, \bibinfo {author} {\bibfnamefont {F.}~\bibnamefont {Vewinger}},\ and\ \bibinfo {author} {\bibfnamefont {M.}~\bibnamefont {Weitz}},\ }\bibfield  {title} {\bibinfo {title} {{Bose–Einstein condensation of photons in an optical microcavity}},\ }\href {https://doi.org/10.1038/nature09567} {\bibfield  {journal} {\bibinfo  {journal} {Nature}\ }\textbf {\bibinfo {volume} {468}},\ \bibinfo {pages} {545} (\bibinfo {year} {2010})}\BibitemShut {NoStop}%
\bibitem [{\citenamefont {Marelic}\ and\ \citenamefont {Nyman}(2015)}]{Nyman15}%
  \BibitemOpen
  \bibfield  {author} {\bibinfo {author} {\bibfnamefont {J.}~\bibnamefont {Marelic}}\ and\ \bibinfo {author} {\bibfnamefont {R.~A.}\ \bibnamefont {Nyman}},\ }\bibfield  {title} {\bibinfo {title} {{Experimental evidence for inhomogeneous pumping and energy-dependent effects in photon Bose-Einstein condensation}},\ }\href {https://doi.org/10.1103/PhysRevA.91.033813} {\bibfield  {journal} {\bibinfo  {journal} {Phys. Rev. A}\ }\textbf {\bibinfo {volume} {91}},\ \bibinfo {pages} {033813} (\bibinfo {year} {2015})}\BibitemShut {NoStop}%
\bibitem [{\citenamefont {Greveling}\ \emph {et~al.}(2018)\citenamefont {Greveling}, \citenamefont {Perrier},\ and\ \citenamefont {van Oosten}}]{Oosten18}%
  \BibitemOpen
  \bibfield  {author} {\bibinfo {author} {\bibfnamefont {S.}~\bibnamefont {Greveling}}, \bibinfo {author} {\bibfnamefont {K.~L.}\ \bibnamefont {Perrier}},\ and\ \bibinfo {author} {\bibfnamefont {D.}~\bibnamefont {van Oosten}},\ }\bibfield  {title} {\bibinfo {title} {{Density distribution of a Bose-Einstein condensate of photons in a dye-filled microcavity}},\ }\href {https://doi.org/10.1103/PhysRevA.98.013810} {\bibfield  {journal} {\bibinfo  {journal} {Phys. Rev. A}\ }\textbf {\bibinfo {volume} {98}},\ \bibinfo {pages} {013810} (\bibinfo {year} {2018})}\BibitemShut {NoStop}%
\bibitem [{\citenamefont {Bloch}\ \emph {et~al.}(2022)\citenamefont {Bloch}, \citenamefont {Carusotto},\ and\ \citenamefont {Wouters}}]{Wouter22}%
  \BibitemOpen
  \bibfield  {author} {\bibinfo {author} {\bibfnamefont {J.}~\bibnamefont {Bloch}}, \bibinfo {author} {\bibfnamefont {I.}~\bibnamefont {Carusotto}},\ and\ \bibinfo {author} {\bibfnamefont {M.}~\bibnamefont {Wouters}},\ }\bibfield  {title} {\bibinfo {title} {{Non-equilibrium Bose–Einstein condensation in photonic systems}},\ }\href {https://doi.org/10.1038/s42254-022-00464-0} {\bibfield  {journal} {\bibinfo  {journal} {Nat Rev Phys}\ }\textbf {\bibinfo {volume} {4}},\ \bibinfo {pages} {470} (\bibinfo {year} {2022})}\BibitemShut {NoStop}%
\bibitem [{\citenamefont {Nyman}\ and\ \citenamefont {Walker}(2018)}]{Nyman-rev18}%
  \BibitemOpen
  \bibfield  {author} {\bibinfo {author} {\bibfnamefont {R.~A.}\ \bibnamefont {Nyman}}\ and\ \bibinfo {author} {\bibfnamefont {B.~T.}\ \bibnamefont {Walker}},\ }\bibfield  {title} {\bibinfo {title} {{Bose-Einstein condensation of photons from the thermodynamic limit to small photon numbers}},\ }\href {https://doi.org/10.1080/09500340.2017.1404655} {\bibfield  {journal} {\bibinfo  {journal} {Journal of Modern Optics}\ }\textbf {\bibinfo {volume} {65}},\ \bibinfo {pages} {754} (\bibinfo {year} {2018})}\BibitemShut {NoStop}%
\bibitem [{\citenamefont {Karkihalli~Umesh}\ \emph {et~al.}(2024)\citenamefont {Karkihalli~Umesh}, \citenamefont {Schulz}, \citenamefont {Schmitt}, \citenamefont {Weitz}, \citenamefont {von Freymann},\ and\ \citenamefont {Vewinger}}]{Kiran24}%
  \BibitemOpen
  \bibfield  {author} {\bibinfo {author} {\bibfnamefont {K.}~\bibnamefont {Karkihalli~Umesh}}, \bibinfo {author} {\bibfnamefont {J.}~\bibnamefont {Schulz}}, \bibinfo {author} {\bibfnamefont {J.}~\bibnamefont {Schmitt}}, \bibinfo {author} {\bibfnamefont {M.}~\bibnamefont {Weitz}}, \bibinfo {author} {\bibfnamefont {G.}~\bibnamefont {von Freymann}},\ and\ \bibinfo {author} {\bibfnamefont {F.}~\bibnamefont {Vewinger}},\ }\bibfield  {title} {\bibinfo {title} {{Dimensional cross-over in a quantum gas of light}},\ }\href {https://doi.org/10.1038/s41567-024-02641-7} {\bibfield  {journal} {\bibinfo  {journal} {Nature Physics}\ }\textbf {\bibinfo {volume} {20}},\ \bibinfo {pages} {1810} (\bibinfo {year} {2024})}\BibitemShut {NoStop}%
\bibitem [{\citenamefont {Walker}\ \emph {et~al.}(2018)\citenamefont {Walker}, \citenamefont {Flatten}, \citenamefont {Hesten}, \citenamefont {Mintert}, \citenamefont {Hunger}, \citenamefont {Trichet}, \citenamefont {Smith},\ and\ \citenamefont {Nyman}}]{Nyman18}%
  \BibitemOpen
  \bibfield  {author} {\bibinfo {author} {\bibfnamefont {B.~T.}\ \bibnamefont {Walker}}, \bibinfo {author} {\bibfnamefont {L.~C.}\ \bibnamefont {Flatten}}, \bibinfo {author} {\bibfnamefont {H.~J.}\ \bibnamefont {Hesten}}, \bibinfo {author} {\bibfnamefont {F.}~\bibnamefont {Mintert}}, \bibinfo {author} {\bibfnamefont {D.}~\bibnamefont {Hunger}}, \bibinfo {author} {\bibfnamefont {A.~A.}\ \bibnamefont {Trichet}}, \bibinfo {author} {\bibfnamefont {J.~M.}\ \bibnamefont {Smith}},\ and\ \bibinfo {author} {\bibfnamefont {R.~A.}\ \bibnamefont {Nyman}},\ }\bibfield  {title} {\bibinfo {title} {{Driven-dissipative non-equilibrium Bose-Einstein condensation of less than ten photons}},\ }\href {https://doi.org/10.1038/s41567-018-0270-1} {\bibfield  {journal} {\bibinfo  {journal} {Nature Physics}\ }\textbf {\bibinfo {volume} {14}},\ \bibinfo {pages} {1173} (\bibinfo {year} {2018})}\BibitemShut {NoStop}%
\bibitem [{\citenamefont {Schmitt}\ \emph {et~al.}(2016)\citenamefont {Schmitt}, \citenamefont {Damm}, \citenamefont {Dung}, \citenamefont {Wahl}, \citenamefont {Vewinger}, \citenamefont {Klaers},\ and\ \citenamefont {Weitz}}]{Schmitt16}%
  \BibitemOpen
  \bibfield  {author} {\bibinfo {author} {\bibfnamefont {J.}~\bibnamefont {Schmitt}}, \bibinfo {author} {\bibfnamefont {T.}~\bibnamefont {Damm}}, \bibinfo {author} {\bibfnamefont {D.}~\bibnamefont {Dung}}, \bibinfo {author} {\bibfnamefont {C.}~\bibnamefont {Wahl}}, \bibinfo {author} {\bibfnamefont {F.}~\bibnamefont {Vewinger}}, \bibinfo {author} {\bibfnamefont {J.}~\bibnamefont {Klaers}},\ and\ \bibinfo {author} {\bibfnamefont {M.}~\bibnamefont {Weitz}},\ }\bibfield  {title} {\bibinfo {title} {{Spontaneous Symmetry Breaking and Phase Coherence of a Photon Bose-Einstein Condensate Coupled to a Reservoir}},\ }\href {https://doi.org/10.1103/PhysRevLett.116.033604} {\bibfield  {journal} {\bibinfo  {journal} {Phys. Rev. Lett.}\ }\textbf {\bibinfo {volume} {116}},\ \bibinfo {pages} {033604} (\bibinfo {year} {2016})}\BibitemShut {NoStop}%
\bibitem [{\citenamefont {Busley}\ \emph {et~al.}(2022)\citenamefont {Busley}, \citenamefont {Miranda}, \citenamefont {Redmann}, \citenamefont {Kurtscheid}, \citenamefont {Umesh}, \citenamefont {Vewinger}, \citenamefont {Weitz},\ and\ \citenamefont {Schmitt}}]{Schmitt22}%
  \BibitemOpen
  \bibfield  {author} {\bibinfo {author} {\bibfnamefont {E.}~\bibnamefont {Busley}}, \bibinfo {author} {\bibfnamefont {L.~E.}\ \bibnamefont {Miranda}}, \bibinfo {author} {\bibfnamefont {A.}~\bibnamefont {Redmann}}, \bibinfo {author} {\bibfnamefont {C.}~\bibnamefont {Kurtscheid}}, \bibinfo {author} {\bibfnamefont {K.~K.}\ \bibnamefont {Umesh}}, \bibinfo {author} {\bibfnamefont {F.}~\bibnamefont {Vewinger}}, \bibinfo {author} {\bibfnamefont {M.}~\bibnamefont {Weitz}},\ and\ \bibinfo {author} {\bibfnamefont {J.}~\bibnamefont {Schmitt}},\ }\bibfield  {title} {\bibinfo {title} {{Compressibility and the equation of state of an optical quantum gas in a box}},\ }\href {https://doi.org/10.1126/science.abm2543} {\bibfield  {journal} {\bibinfo  {journal} {Science}\ }\textbf {\bibinfo {volume} {375}},\ \bibinfo {pages} {1403} (\bibinfo {year} {2022})}\BibitemShut {NoStop}%
\bibitem [{\citenamefont {\"Ozt\"urk}\ \emph {et~al.}(2023)\citenamefont {\"Ozt\"urk}, \citenamefont {Vewinger}, \citenamefont {Weitz},\ and\ \citenamefont {Schmitt}}]{Schmitt23}%
  \BibitemOpen
  \bibfield  {author} {\bibinfo {author} {\bibfnamefont {F.~E.}\ \bibnamefont {\"Ozt\"urk}}, \bibinfo {author} {\bibfnamefont {F.}~\bibnamefont {Vewinger}}, \bibinfo {author} {\bibfnamefont {M.}~\bibnamefont {Weitz}},\ and\ \bibinfo {author} {\bibfnamefont {J.}~\bibnamefont {Schmitt}},\ }\bibfield  {title} {\bibinfo {title} {{Fluctuation-Dissipation Relation for a Bose-Einstein Condensate of Photons}},\ }\href {https://doi.org/10.1103/PhysRevLett.130.033602} {\bibfield  {journal} {\bibinfo  {journal} {Phys. Rev. Lett.}\ }\textbf {\bibinfo {volume} {130}},\ \bibinfo {pages} {033602} (\bibinfo {year} {2023})}\BibitemShut {NoStop}%
\bibitem [{\citenamefont {Öztürk}\ \emph {et~al.}(2021)\citenamefont {Öztürk}, \citenamefont {Lappe}, \citenamefont {Hellmann}, \citenamefont {Schmitt}, \citenamefont {Klaers}, \citenamefont {Vewinger}, \citenamefont {Kroha},\ and\ \citenamefont {Weitz}}]{Ozturk21}%
  \BibitemOpen
  \bibfield  {author} {\bibinfo {author} {\bibfnamefont {F.~E.}\ \bibnamefont {Öztürk}}, \bibinfo {author} {\bibfnamefont {T.}~\bibnamefont {Lappe}}, \bibinfo {author} {\bibfnamefont {G.}~\bibnamefont {Hellmann}}, \bibinfo {author} {\bibfnamefont {J.}~\bibnamefont {Schmitt}}, \bibinfo {author} {\bibfnamefont {J.}~\bibnamefont {Klaers}}, \bibinfo {author} {\bibfnamefont {F.}~\bibnamefont {Vewinger}}, \bibinfo {author} {\bibfnamefont {J.}~\bibnamefont {Kroha}},\ and\ \bibinfo {author} {\bibfnamefont {M.}~\bibnamefont {Weitz}},\ }\bibfield  {title} {\bibinfo {title} {{Observation of a non-Hermitian phase transition in an optical quantum gas}},\ }\href {https://doi.org/10.1126/science.abe9869} {\bibfield  {journal} {\bibinfo  {journal} {Science}\ }\textbf {\bibinfo {volume} {372}},\ \bibinfo {pages} {88} (\bibinfo {year} {2021})}\BibitemShut {NoStop}%
\bibitem [{\citenamefont {Pieczarka}\ \emph {et~al.}(2024)\citenamefont {Pieczarka}, \citenamefont {G{\k{e}}bski}, \citenamefont {Piasecka}, \citenamefont {Lott}, \citenamefont {Pelster}, \citenamefont {Wasiak},\ and\ \citenamefont {Czyszanowski}}]{Pieczarka24}%
  \BibitemOpen
  \bibfield  {author} {\bibinfo {author} {\bibfnamefont {M.}~\bibnamefont {Pieczarka}}, \bibinfo {author} {\bibfnamefont {M.}~\bibnamefont {G{\k{e}}bski}}, \bibinfo {author} {\bibfnamefont {A.~N.}\ \bibnamefont {Piasecka}}, \bibinfo {author} {\bibfnamefont {J.~A.}\ \bibnamefont {Lott}}, \bibinfo {author} {\bibfnamefont {A.}~\bibnamefont {Pelster}}, \bibinfo {author} {\bibfnamefont {M.}~\bibnamefont {Wasiak}},\ and\ \bibinfo {author} {\bibfnamefont {T.}~\bibnamefont {Czyszanowski}},\ }\bibfield  {title} {\bibinfo {title} {{Bose--Einstein condensation of photons in a vertical-cavity surface-emitting laser}},\ }\href {https://doi.org/10.1038/s41566-024-01478-z} {\bibfield  {journal} {\bibinfo  {journal} {Nature Photonics}\ }\textbf {\bibinfo {volume} {18}},\ \bibinfo {pages} {1090} (\bibinfo {year} {2024})}\BibitemShut {NoStop}%
\bibitem [{\citenamefont {Schmitt}\ \emph {et~al.}(2014)\citenamefont {Schmitt}, \citenamefont {Damm}, \citenamefont {Dung}, \citenamefont {Vewinger}, \citenamefont {Klaers},\ and\ \citenamefont {Weitz}}]{Schmitt14}%
  \BibitemOpen
  \bibfield  {author} {\bibinfo {author} {\bibfnamefont {J.}~\bibnamefont {Schmitt}}, \bibinfo {author} {\bibfnamefont {T.}~\bibnamefont {Damm}}, \bibinfo {author} {\bibfnamefont {D.}~\bibnamefont {Dung}}, \bibinfo {author} {\bibfnamefont {F.}~\bibnamefont {Vewinger}}, \bibinfo {author} {\bibfnamefont {J.}~\bibnamefont {Klaers}},\ and\ \bibinfo {author} {\bibfnamefont {M.}~\bibnamefont {Weitz}},\ }\bibfield  {title} {\bibinfo {title} {{Observation of Grand-Canonical Number Statistics in a Photon Bose-Einstein Condensate}},\ }\href {https://doi.org/10.1103/PhysRevLett.112.030401} {\bibfield  {journal} {\bibinfo  {journal} {Phys. Rev. Lett.}\ }\textbf {\bibinfo {volume} {112}},\ \bibinfo {pages} {030401} (\bibinfo {year} {2014})}\BibitemShut {NoStop}%
\bibitem [{\citenamefont {Marelic}\ \emph {et~al.}(2016)\citenamefont {Marelic}, \citenamefont {Walker},\ and\ \citenamefont {Nyman}}]{Nyman16}%
  \BibitemOpen
  \bibfield  {author} {\bibinfo {author} {\bibfnamefont {J.}~\bibnamefont {Marelic}}, \bibinfo {author} {\bibfnamefont {B.~T.}\ \bibnamefont {Walker}},\ and\ \bibinfo {author} {\bibfnamefont {R.~A.}\ \bibnamefont {Nyman}},\ }\bibfield  {title} {\bibinfo {title} {{Phase-space views into dye-microcavity thermalized and condensed photons}},\ }\href {https://doi.org/10.1103/PhysRevA.94.063812} {\bibfield  {journal} {\bibinfo  {journal} {Phys. Rev. A}\ }\textbf {\bibinfo {volume} {94}},\ \bibinfo {pages} {063812} (\bibinfo {year} {2016})}\BibitemShut {NoStop}%
\bibitem [{\citenamefont {Tang}\ \emph {et~al.}(2024)\citenamefont {Tang}, \citenamefont {Dhar}, \citenamefont {Oulton}, \citenamefont {Nyman},\ and\ \citenamefont {Mintert}}]{Nyman24}%
  \BibitemOpen
  \bibfield  {author} {\bibinfo {author} {\bibfnamefont {Y.}~\bibnamefont {Tang}}, \bibinfo {author} {\bibfnamefont {H.~S.}\ \bibnamefont {Dhar}}, \bibinfo {author} {\bibfnamefont {R.~F.}\ \bibnamefont {Oulton}}, \bibinfo {author} {\bibfnamefont {R.~A.}\ \bibnamefont {Nyman}},\ and\ \bibinfo {author} {\bibfnamefont {F.}~\bibnamefont {Mintert}},\ }\bibfield  {title} {\bibinfo {title} {{Breakdown of Temporal Coherence in Photon Condensates}},\ }\href {https://doi.org/10.1103/PhysRevLett.132.173601} {\bibfield  {journal} {\bibinfo  {journal} {Phys. Rev. Lett.}\ }\textbf {\bibinfo {volume} {132}},\ \bibinfo {pages} {173601} (\bibinfo {year} {2024})}\BibitemShut {NoStop}%
\bibitem [{\citenamefont {Ozturk}\ \emph {et~al.}(2019)\citenamefont {Ozturk}, \citenamefont {Lappe}, \citenamefont {Hellmann}, \citenamefont {Schmitt}, \citenamefont {Klaers}, \citenamefont {Vewinger}, \citenamefont {Kroha},\ and\ \citenamefont {Weitz}}]{TimBode19}%
  \BibitemOpen
  \bibfield  {author} {\bibinfo {author} {\bibfnamefont {F.~E.}\ \bibnamefont {Ozturk}}, \bibinfo {author} {\bibfnamefont {T.}~\bibnamefont {Lappe}}, \bibinfo {author} {\bibfnamefont {G.}~\bibnamefont {Hellmann}}, \bibinfo {author} {\bibfnamefont {J.}~\bibnamefont {Schmitt}}, \bibinfo {author} {\bibfnamefont {J.}~\bibnamefont {Klaers}}, \bibinfo {author} {\bibfnamefont {F.}~\bibnamefont {Vewinger}}, \bibinfo {author} {\bibfnamefont {J.}~\bibnamefont {Kroha}},\ and\ \bibinfo {author} {\bibfnamefont {M.}~\bibnamefont {Weitz}},\ }\bibfield  {title} {\bibinfo {title} {{Fluctuation dynamics of an open photon Bose-Einstein condensate}},\ }\href {https://doi.org/10.1103/PhysRevA.100.043803} {\bibfield  {journal} {\bibinfo  {journal} {Phys. Rev. A}\ }\textbf {\bibinfo {volume} {100}},\ \bibinfo {pages} {043803} (\bibinfo {year} {2019})}\BibitemShut {NoStop}%
\bibitem [{\citenamefont {Gladilin}\ and\ \citenamefont {Wouters}(2020{\natexlab{a}})}]{Wouter20}%
  \BibitemOpen
  \bibfield  {author} {\bibinfo {author} {\bibfnamefont {V.~N.}\ \bibnamefont {Gladilin}}\ and\ \bibinfo {author} {\bibfnamefont {M.}~\bibnamefont {Wouters}},\ }\bibfield  {title} {\bibinfo {title} {{Classical field model for arrays of photon condensates}},\ }\href {https://doi.org/10.1103/PhysRevA.101.043814} {\bibfield  {journal} {\bibinfo  {journal} {Phys. Rev. A}\ }\textbf {\bibinfo {volume} {101}},\ \bibinfo {pages} {043814} (\bibinfo {year} {2020}{\natexlab{a}})}\BibitemShut {NoStop}%
\bibitem [{\citenamefont {Gladilin}\ and\ \citenamefont {Wouters}(2020{\natexlab{b}})}]{Gladlin20}%
  \BibitemOpen
  \bibfield  {author} {\bibinfo {author} {\bibfnamefont {V.~N.}\ \bibnamefont {Gladilin}}\ and\ \bibinfo {author} {\bibfnamefont {M.}~\bibnamefont {Wouters}},\ }\bibfield  {title} {\bibinfo {title} {{Vortices in Nonequilibrium Photon Condensates}},\ }\href {https://doi.org/10.1103/PhysRevLett.125.215301} {\bibfield  {journal} {\bibinfo  {journal} {Phys. Rev. Lett.}\ }\textbf {\bibinfo {volume} {125}},\ \bibinfo {pages} {215301} (\bibinfo {year} {2020}{\natexlab{b}})}\BibitemShut {NoStop}%
\bibitem [{\citenamefont {Kirton}\ and\ \citenamefont {Keeling}(2013)}]{Keeling13}%
  \BibitemOpen
  \bibfield  {author} {\bibinfo {author} {\bibfnamefont {P.}~\bibnamefont {Kirton}}\ and\ \bibinfo {author} {\bibfnamefont {J.}~\bibnamefont {Keeling}},\ }\bibfield  {title} {\bibinfo {title} {{Nonequilibrium Model of Photon Condensation}},\ }\href {https://doi.org/10.1103/PhysRevLett.111.100404} {\bibfield  {journal} {\bibinfo  {journal} {Phys. Rev. Lett.}\ }\textbf {\bibinfo {volume} {111}},\ \bibinfo {pages} {100404} (\bibinfo {year} {2013})}\BibitemShut {NoStop}%
\bibitem [{\citenamefont {Kirton}\ and\ \citenamefont {Keeling}(2015)}]{Keeling15}%
  \BibitemOpen
  \bibfield  {author} {\bibinfo {author} {\bibfnamefont {P.}~\bibnamefont {Kirton}}\ and\ \bibinfo {author} {\bibfnamefont {J.}~\bibnamefont {Keeling}},\ }\bibfield  {title} {\bibinfo {title} {{Thermalization and breakdown of thermalization in photon condensates}},\ }\href {https://doi.org/10.1103/PhysRevA.91.033826} {\bibfield  {journal} {\bibinfo  {journal} {Phys. Rev. A}\ }\textbf {\bibinfo {volume} {91}},\ \bibinfo {pages} {033826} (\bibinfo {year} {2015})}\BibitemShut {NoStop}%
\bibitem [{\citenamefont {Radonjić}\ \emph {et~al.}(2018)\citenamefont {Radonjić}, \citenamefont {Kopylov}, \citenamefont {Balaž},\ and\ \citenamefont {Pelster}}]{Pelster}%
  \BibitemOpen
  \bibfield  {author} {\bibinfo {author} {\bibfnamefont {M.}~\bibnamefont {Radonjić}}, \bibinfo {author} {\bibfnamefont {W.}~\bibnamefont {Kopylov}}, \bibinfo {author} {\bibfnamefont {A.}~\bibnamefont {Balaž}},\ and\ \bibinfo {author} {\bibfnamefont {A.}~\bibnamefont {Pelster}},\ }\bibfield  {title} {\bibinfo {title} {{Interplay of coherent and dissipative dynamics in condensates of light}},\ }\href {https://doi.org/10.1088/1367-2630/aac2a6} {\bibfield  {journal} {\bibinfo  {journal} {New Journal of Physics}\ }\textbf {\bibinfo {volume} {20}},\ \bibinfo {pages} {055014} (\bibinfo {year} {2018})}\BibitemShut {NoStop}%
\bibitem [{\citenamefont {Walker}\ \emph {et~al.}(2019)\citenamefont {Walker}, \citenamefont {Hesten}, \citenamefont {Dhar}, \citenamefont {Nyman},\ and\ \citenamefont {Mintert}}]{Nyman19}%
  \BibitemOpen
  \bibfield  {author} {\bibinfo {author} {\bibfnamefont {B.~T.}\ \bibnamefont {Walker}}, \bibinfo {author} {\bibfnamefont {H.~J.}\ \bibnamefont {Hesten}}, \bibinfo {author} {\bibfnamefont {H.~S.}\ \bibnamefont {Dhar}}, \bibinfo {author} {\bibfnamefont {R.~A.}\ \bibnamefont {Nyman}},\ and\ \bibinfo {author} {\bibfnamefont {F.}~\bibnamefont {Mintert}},\ }\bibfield  {title} {\bibinfo {title} {{Noncritical Slowing Down of Photonic Condensation}},\ }\href {https://doi.org/10.1103/PhysRevLett.123.203602} {\bibfield  {journal} {\bibinfo  {journal} {Phys. Rev. Lett.}\ }\textbf {\bibinfo {volume} {123}},\ \bibinfo {pages} {203602} (\bibinfo {year} {2019})}\BibitemShut {NoStop}%
\bibitem [{\citenamefont {Bode}\ \emph {et~al.}(2024)\citenamefont {Bode}, \citenamefont {Kajan}, \citenamefont {Meirinhos},\ and\ \citenamefont {Kroha}}]{TimBode24}%
  \BibitemOpen
  \bibfield  {author} {\bibinfo {author} {\bibfnamefont {T.}~\bibnamefont {Bode}}, \bibinfo {author} {\bibfnamefont {M.}~\bibnamefont {Kajan}}, \bibinfo {author} {\bibfnamefont {F.}~\bibnamefont {Meirinhos}},\ and\ \bibinfo {author} {\bibfnamefont {J.}~\bibnamefont {Kroha}},\ }\bibfield  {title} {\bibinfo {title} {{Non-Markovian dynamics of open quantum systems via auxiliary particles with exact operator constraint}},\ }\href {https://doi.org/10.1103/PhysRevResearch.6.013220} {\bibfield  {journal} {\bibinfo  {journal} {Phys. Rev. Res.}\ }\textbf {\bibinfo {volume} {6}},\ \bibinfo {pages} {013220} (\bibinfo {year} {2024})}\BibitemShut {NoStop}%
\bibitem [{\citenamefont {Fowler-Wright}\ \emph {et~al.}(2022)\citenamefont {Fowler-Wright}, \citenamefont {Lovett},\ and\ \citenamefont {Keeling}}]{Keeling-symmetry-breaking}%
  \BibitemOpen
  \bibfield  {author} {\bibinfo {author} {\bibfnamefont {P.}~\bibnamefont {Fowler-Wright}}, \bibinfo {author} {\bibfnamefont {B.~W.}\ \bibnamefont {Lovett}},\ and\ \bibinfo {author} {\bibfnamefont {J.}~\bibnamefont {Keeling}},\ }\bibfield  {title} {\bibinfo {title} {{Efficient Many-Body Non-Markovian Dynamics of Organic Polaritons}},\ }\href {https://doi.org/10.1103/PhysRevLett.129.173001} {\bibfield  {journal} {\bibinfo  {journal} {Phys. Rev. Lett.}\ }\textbf {\bibinfo {volume} {129}},\ \bibinfo {pages} {173001} (\bibinfo {year} {2022})}\BibitemShut {NoStop}%
\bibitem [{\citenamefont {Strogatz}(2001)}]{Strogatz-book}%
  \BibitemOpen
  \bibfield  {author} {\bibinfo {author} {\bibfnamefont {S.~H.}\ \bibnamefont {Strogatz}},\ }\href@noop {} {\emph {\bibinfo {title} {{Nonlinear dynamics and chaos: with applications to physics, biology, chemistry, and engineering (studies in nonlinearity)}}}},\ Vol.~\bibinfo {volume} {1}\ (\bibinfo  {publisher} {Westview Press},\ \bibinfo {year} {2001})\BibitemShut {NoStop}%
\bibitem [{\citenamefont {Strogatz}\ and\ \citenamefont {Westervelt}(1989)}]{Strogatz89}%
  \BibitemOpen
  \bibfield  {author} {\bibinfo {author} {\bibfnamefont {S.~H.}\ \bibnamefont {Strogatz}}\ and\ \bibinfo {author} {\bibfnamefont {R.~M.}\ \bibnamefont {Westervelt}},\ }\bibfield  {title} {\bibinfo {title} {{Predicted power laws for delayed switching of charge-density waves}},\ }\href {https://doi.org/10.1103/PhysRevB.40.10501} {\bibfield  {journal} {\bibinfo  {journal} {Phys. Rev. B}\ }\textbf {\bibinfo {volume} {40}},\ \bibinfo {pages} {10501} (\bibinfo {year} {1989})}\BibitemShut {NoStop}%
\bibitem [{\citenamefont {Koch}\ \emph {et~al.}(2024)\citenamefont {Koch}, \citenamefont {Nandan}, \citenamefont {Ramesan}, \citenamefont {Tyukin}, \citenamefont {Gorban},\ and\ \citenamefont {Koseska}}]{Koch24}%
  \BibitemOpen
  \bibfield  {author} {\bibinfo {author} {\bibfnamefont {D.}~\bibnamefont {Koch}}, \bibinfo {author} {\bibfnamefont {A.}~\bibnamefont {Nandan}}, \bibinfo {author} {\bibfnamefont {G.}~\bibnamefont {Ramesan}}, \bibinfo {author} {\bibfnamefont {I.}~\bibnamefont {Tyukin}}, \bibinfo {author} {\bibfnamefont {A.}~\bibnamefont {Gorban}},\ and\ \bibinfo {author} {\bibfnamefont {A.}~\bibnamefont {Koseska}},\ }\bibfield  {title} {\bibinfo {title} {{Ghost Channels and Ghost Cycles Guiding Long Transients in Dynamical Systems}},\ }\href {https://doi.org/10.1103/PhysRevLett.133.047202} {\bibfield  {journal} {\bibinfo  {journal} {Phys. Rev. Lett.}\ }\textbf {\bibinfo {volume} {133}},\ \bibinfo {pages} {047202} (\bibinfo {year} {2024})}\BibitemShut {NoStop}%
\bibitem [{\citenamefont {Condon}(1928)}]{Condon1928}%
  \BibitemOpen
  \bibfield  {author} {\bibinfo {author} {\bibfnamefont {E.~U.}\ \bibnamefont {Condon}},\ }\bibfield  {title} {\bibinfo {title} {{Nuclear Motions Associated with Electron Transitions in Diatomic Molecules}},\ }\href {https://doi.org/10.1103/PhysRev.32.858} {\bibfield  {journal} {\bibinfo  {journal} {Phys. Rev.}\ }\textbf {\bibinfo {volume} {32}},\ \bibinfo {pages} {858} (\bibinfo {year} {1928})}\BibitemShut {NoStop}%
\bibitem [{\citenamefont {Breuer}\ and\ \citenamefont {Petruccione}(2007)}]{Petruccione_book}%
  \BibitemOpen
  \bibfield  {author} {\bibinfo {author} {\bibfnamefont {H.-P.}\ \bibnamefont {Breuer}}\ and\ \bibinfo {author} {\bibfnamefont {F.}~\bibnamefont {Petruccione}},\ }\href {https://doi.org/10.1093/acprof:oso/9780199213900.001.0001} {\emph {\bibinfo {title} {{The Theory of Open Quantum Systems}}}}\ (\bibinfo  {publisher} {Oxford University Press},\ \bibinfo {year} {2007})\BibitemShut {NoStop}%
\bibitem [{\citenamefont {Moilanen}\ \emph {et~al.}(2022)\citenamefont {Moilanen}, \citenamefont {Arnard\'ottir}, \citenamefont {Keeling},\ and\ \citenamefont {T\"orm\"a}}]{Keeling22}%
  \BibitemOpen
  \bibfield  {author} {\bibinfo {author} {\bibfnamefont {A.~J.}\ \bibnamefont {Moilanen}}, \bibinfo {author} {\bibfnamefont {K.~B.}\ \bibnamefont {Arnard\'ottir}}, \bibinfo {author} {\bibfnamefont {J.}~\bibnamefont {Keeling}},\ and\ \bibinfo {author} {\bibfnamefont {P.}~\bibnamefont {T\"orm\"a}},\ }\bibfield  {title} {\bibinfo {title} {{Mode switching dynamics in organic polariton lasing}},\ }\href {https://doi.org/10.1103/PhysRevB.106.195403} {\bibfield  {journal} {\bibinfo  {journal} {Phys. Rev. B}\ }\textbf {\bibinfo {volume} {106}},\ \bibinfo {pages} {195403} (\bibinfo {year} {2022})}\BibitemShut {NoStop}%
\bibitem [{\citenamefont {Keeling}\ and\ \citenamefont {Kéna-Cohen}(2020)}]{Keeling-polariton20}%
  \BibitemOpen
  \bibfield  {author} {\bibinfo {author} {\bibfnamefont {J.}~\bibnamefont {Keeling}}\ and\ \bibinfo {author} {\bibfnamefont {S.}~\bibnamefont {Kéna-Cohen}},\ }\bibfield  {title} {\bibinfo {title} {{Bose–Einstein Condensation of Exciton-Polaritons in Organic Microcavities}},\ }\href {https://doi.org/https://doi.org/10.1146/annurev-physchem-010920-102509} {\bibfield  {journal} {\bibinfo  {journal} {Annual Review of Physical Chemistry}\ }\textbf {\bibinfo {volume} {71}},\ \bibinfo {pages} {435} (\bibinfo {year} {2020})}\BibitemShut {NoStop}%
\bibitem [{\citenamefont {Carre\~no}\ \emph {et~al.}(2016)\citenamefont {Carre\~no}, \citenamefont {S\'anchez Mu\~noz}, \citenamefont {del Valle},\ and\ \citenamefont {Laussy}}]{Laussy16}%
  \BibitemOpen
  \bibfield  {author} {\bibinfo {author} {\bibfnamefont {J.~C.~L.}\ \bibnamefont {Carre\~no}}, \bibinfo {author} {\bibfnamefont {C.}~\bibnamefont {S\'anchez Mu\~noz}}, \bibinfo {author} {\bibfnamefont {E.}~\bibnamefont {del Valle}},\ and\ \bibinfo {author} {\bibfnamefont {F.~P.}\ \bibnamefont {Laussy}},\ }\bibfield  {title} {\bibinfo {title} {{Excitation with quantum light. II. Exciting a two-level system}},\ }\href {https://doi.org/10.1103/PhysRevA.94.063826} {\bibfield  {journal} {\bibinfo  {journal} {Phys. Rev. A}\ }\textbf {\bibinfo {volume} {94}},\ \bibinfo {pages} {063826} (\bibinfo {year} {2016})}\BibitemShut {NoStop}%
\bibitem [{\citenamefont {Schmitt}\ \emph {et~al.}(2024)\citenamefont {Schmitt}, \citenamefont {Weitz},\ and\ \citenamefont {Klaers}}]{schmitt_absorption_2024}%
  \BibitemOpen
  \bibfield  {author} {\bibinfo {author} {\bibfnamefont {J.}~\bibnamefont {Schmitt}}, \bibinfo {author} {\bibfnamefont {M.}~\bibnamefont {Weitz}},\ and\ \bibinfo {author} {\bibfnamefont {J.}~\bibnamefont {Klaers}},\ }\bibfield  {title} {\bibinfo {title} {Absorption and emission spectral data of room- temperature rhodamine 6g dye solution and some typical dye microcavity parameters},\ }\bibfield  {journal} {\bibinfo  {journal} {Zenodo,}\ }\href {https://doi.org/10.5281/zenodo.10852936} {10.5281/zenodo.10852936} (\bibinfo {year} {2024})\BibitemShut {NoStop}%
\bibitem [{Sup()}]{Supplement}%
  \BibitemOpen
  \href@noop {} {}\bibinfo {note} {See the Supplemental Material below for the detailed calculation of fixed points and their linear stability analysis, and representative dynamics at lower pumping with fewer photon number.}\BibitemShut {Stop}%
\bibitem [{\citenamefont {Moeckel}\ and\ \citenamefont {Kehrein}(2008)}]{Moeckel08}%
  \BibitemOpen
  \bibfield  {author} {\bibinfo {author} {\bibfnamefont {M.}~\bibnamefont {Moeckel}}\ and\ \bibinfo {author} {\bibfnamefont {S.}~\bibnamefont {Kehrein}},\ }\bibfield  {title} {\bibinfo {title} {{Interaction Quench in the Hubbard Model}},\ }\href {https://doi.org/10.1103/PhysRevLett.100.175702} {\bibfield  {journal} {\bibinfo  {journal} {Phys. Rev. Lett.}\ }\textbf {\bibinfo {volume} {100}},\ \bibinfo {pages} {175702} (\bibinfo {year} {2008})}\BibitemShut {NoStop}%
\bibitem [{\citenamefont {Kollar}\ \emph {et~al.}(2011)\citenamefont {Kollar}, \citenamefont {Wolf},\ and\ \citenamefont {Eckstein}}]{Eckstein11}%
  \BibitemOpen
  \bibfield  {author} {\bibinfo {author} {\bibfnamefont {M.}~\bibnamefont {Kollar}}, \bibinfo {author} {\bibfnamefont {F.~A.}\ \bibnamefont {Wolf}},\ and\ \bibinfo {author} {\bibfnamefont {M.}~\bibnamefont {Eckstein}},\ }\bibfield  {title} {\bibinfo {title} {{Generalized Gibbs ensemble prediction of prethermalization plateaus and their relation to nonthermal steady states in integrable systems}},\ }\href {https://doi.org/10.1103/PhysRevB.84.054304} {\bibfield  {journal} {\bibinfo  {journal} {Phys. Rev. B}\ }\textbf {\bibinfo {volume} {84}},\ \bibinfo {pages} {054304} (\bibinfo {year} {2011})}\BibitemShut {NoStop}%
\bibitem [{\citenamefont {Bertini}\ \emph {et~al.}(2015)\citenamefont {Bertini}, \citenamefont {Essler}, \citenamefont {Groha},\ and\ \citenamefont {Robinson}}]{Essler15}%
  \BibitemOpen
  \bibfield  {author} {\bibinfo {author} {\bibfnamefont {B.}~\bibnamefont {Bertini}}, \bibinfo {author} {\bibfnamefont {F.~H.~L.}\ \bibnamefont {Essler}}, \bibinfo {author} {\bibfnamefont {S.}~\bibnamefont {Groha}},\ and\ \bibinfo {author} {\bibfnamefont {N.~J.}\ \bibnamefont {Robinson}},\ }\bibfield  {title} {\bibinfo {title} {{Prethermalization and Thermalization in Models with Weak Integrability Breaking}},\ }\href {https://doi.org/10.1103/PhysRevLett.115.180601} {\bibfield  {journal} {\bibinfo  {journal} {Phys. Rev. Lett.}\ }\textbf {\bibinfo {volume} {115}},\ \bibinfo {pages} {180601} (\bibinfo {year} {2015})}\BibitemShut {NoStop}%
\bibitem [{\citenamefont {Mallayya}\ \emph {et~al.}(2019)\citenamefont {Mallayya}, \citenamefont {Rigol},\ and\ \citenamefont {De~Roeck}}]{Rigol19}%
  \BibitemOpen
  \bibfield  {author} {\bibinfo {author} {\bibfnamefont {K.}~\bibnamefont {Mallayya}}, \bibinfo {author} {\bibfnamefont {M.}~\bibnamefont {Rigol}},\ and\ \bibinfo {author} {\bibfnamefont {W.}~\bibnamefont {De~Roeck}},\ }\bibfield  {title} {\bibinfo {title} {{Prethermalization and Thermalization in Isolated Quantum Systems}},\ }\href {https://doi.org/10.1103/PhysRevX.9.021027} {\bibfield  {journal} {\bibinfo  {journal} {Phys. Rev. X}\ }\textbf {\bibinfo {volume} {9}},\ \bibinfo {pages} {021027} (\bibinfo {year} {2019})}\BibitemShut {NoStop}%
\bibitem [{\citenamefont {Ray}\ \emph {et~al.}(2020)\citenamefont {Ray}, \citenamefont {Anglin},\ and\ \citenamefont {Vardi}}]{Ray20}%
  \BibitemOpen
  \bibfield  {author} {\bibinfo {author} {\bibfnamefont {S.}~\bibnamefont {Ray}}, \bibinfo {author} {\bibfnamefont {J.~R.}\ \bibnamefont {Anglin}},\ and\ \bibinfo {author} {\bibfnamefont {A.}~\bibnamefont {Vardi}},\ }\bibfield  {title} {\bibinfo {title} {{Prethermalization with negative specific heat}},\ }\href {https://doi.org/10.1103/PhysRevE.102.052107} {\bibfield  {journal} {\bibinfo  {journal} {Phys. Rev. E}\ }\textbf {\bibinfo {volume} {102}},\ \bibinfo {pages} {052107} (\bibinfo {year} {2020})}\BibitemShut {NoStop}%
\bibitem [{\citenamefont {Hanai}\ \emph {et~al.}(2019)\citenamefont {Hanai}, \citenamefont {Edelman}, \citenamefont {Ohashi},\ and\ \citenamefont {Littlewood}}]{Littlewood19}%
  \BibitemOpen
  \bibfield  {author} {\bibinfo {author} {\bibfnamefont {R.}~\bibnamefont {Hanai}}, \bibinfo {author} {\bibfnamefont {A.}~\bibnamefont {Edelman}}, \bibinfo {author} {\bibfnamefont {Y.}~\bibnamefont {Ohashi}},\ and\ \bibinfo {author} {\bibfnamefont {P.~B.}\ \bibnamefont {Littlewood}},\ }\bibfield  {title} {\bibinfo {title} {{Non-Hermitian Phase Transition from a Polariton Bose-Einstein Condensate to a Photon Laser}},\ }\href {https://doi.org/10.1103/PhysRevLett.122.185301} {\bibfield  {journal} {\bibinfo  {journal} {Phys. Rev. Lett.}\ }\textbf {\bibinfo {volume} {122}},\ \bibinfo {pages} {185301} (\bibinfo {year} {2019})}\BibitemShut {NoStop}%
\bibitem [{\citenamefont {Bezvershenko}\ \emph {et~al.}(2021)\citenamefont {Bezvershenko}, \citenamefont {Halati}, \citenamefont {Sheikhan}, \citenamefont {Kollath},\ and\ \citenamefont {Rosch}}]{Kollath}%
  \BibitemOpen
  \bibfield  {author} {\bibinfo {author} {\bibfnamefont {A.~V.}\ \bibnamefont {Bezvershenko}}, \bibinfo {author} {\bibfnamefont {C.-M.}\ \bibnamefont {Halati}}, \bibinfo {author} {\bibfnamefont {A.}~\bibnamefont {Sheikhan}}, \bibinfo {author} {\bibfnamefont {C.}~\bibnamefont {Kollath}},\ and\ \bibinfo {author} {\bibfnamefont {A.}~\bibnamefont {Rosch}},\ }\bibfield  {title} {\bibinfo {title} {{Dicke Transition in Open Many-Body Systems Determined by Fluctuation Effects}},\ }\href {https://doi.org/10.1103/PhysRevLett.127.173606} {\bibfield  {journal} {\bibinfo  {journal} {Phys. Rev. Lett.}\ }\textbf {\bibinfo {volume} {127}},\ \bibinfo {pages} {173606} (\bibinfo {year} {2021})}\BibitemShut {NoStop}%
\bibitem [{\citenamefont {Wu}\ \emph {et~al.}(2024)\citenamefont {Wu}, \citenamefont {Ray},\ and\ \citenamefont {Kroha}}]{Ray24}%
  \BibitemOpen
  \bibfield  {author} {\bibinfo {author} {\bibfnamefont {T.}~\bibnamefont {Wu}}, \bibinfo {author} {\bibfnamefont {S.}~\bibnamefont {Ray}},\ and\ \bibinfo {author} {\bibfnamefont {J.}~\bibnamefont {Kroha}},\ }\bibfield  {title} {\bibinfo {title} {{Temporal Bistability in the Dissipative Dicke-Bose-Hubbard System}},\ }\href {https://doi.org/https://doi.org/10.1002/andp.202300505} {\bibfield  {journal} {\bibinfo  {journal} {Annalen der Physik}\ }\textbf {\bibinfo {volume} {536}},\ \bibinfo {pages} {2300505} (\bibinfo {year} {2024})}\BibitemShut {NoStop}%
\bibitem [{\citenamefont {Abouelela}\ \emph {et~al.}(2025)\citenamefont {Abouelela}, \citenamefont {Turaev}, \citenamefont {Kramer}, \citenamefont {Janning}, \citenamefont {Kajan}, \citenamefont {Ray},\ and\ \citenamefont {Kroha}}]{Zenodo-data_2025}%
  \BibitemOpen
  \bibfield  {author} {\bibinfo {author} {\bibfnamefont {A.}~\bibnamefont {Abouelela}}, \bibinfo {author} {\bibfnamefont {M.}~\bibnamefont {Turaev}}, \bibinfo {author} {\bibfnamefont {R.}~\bibnamefont {Kramer}}, \bibinfo {author} {\bibfnamefont {M.}~\bibnamefont {Janning}}, \bibinfo {author} {\bibfnamefont {M.}~\bibnamefont {Kajan}}, \bibinfo {author} {\bibfnamefont {S.}~\bibnamefont {Ray}},\ and\ \bibinfo {author} {\bibfnamefont {J.}~\bibnamefont {Kroha}},\ }\bibfield  {title} {\bibinfo {title} {Figure data for ``stabilizing open photon condensates by ghost-attractor dynamics"},\ }\bibfield  {journal} {\bibinfo  {journal} {Zenodo,}\ }\href {https://doi.org/10.5281/zenodo.15830145} {10.5281/zenodo.15830145} (\bibinfo {year} {2025})\BibitemShut {NoStop}%
\end{thebibliography}%

\appendix

\renewcommand{\thefigure}{A\arabic{figure}}
\setcounter{figure}{0}
\renewcommand{\theequation}{A\arabic{equation}}
\setcounter{equation}{0}

\section{End Matter}
Here we present additional results on the condensate and fluctuation dynamics in different parameter regimes.

\begin{figure}[b]
\includegraphics[width=0.82\columnwidth]{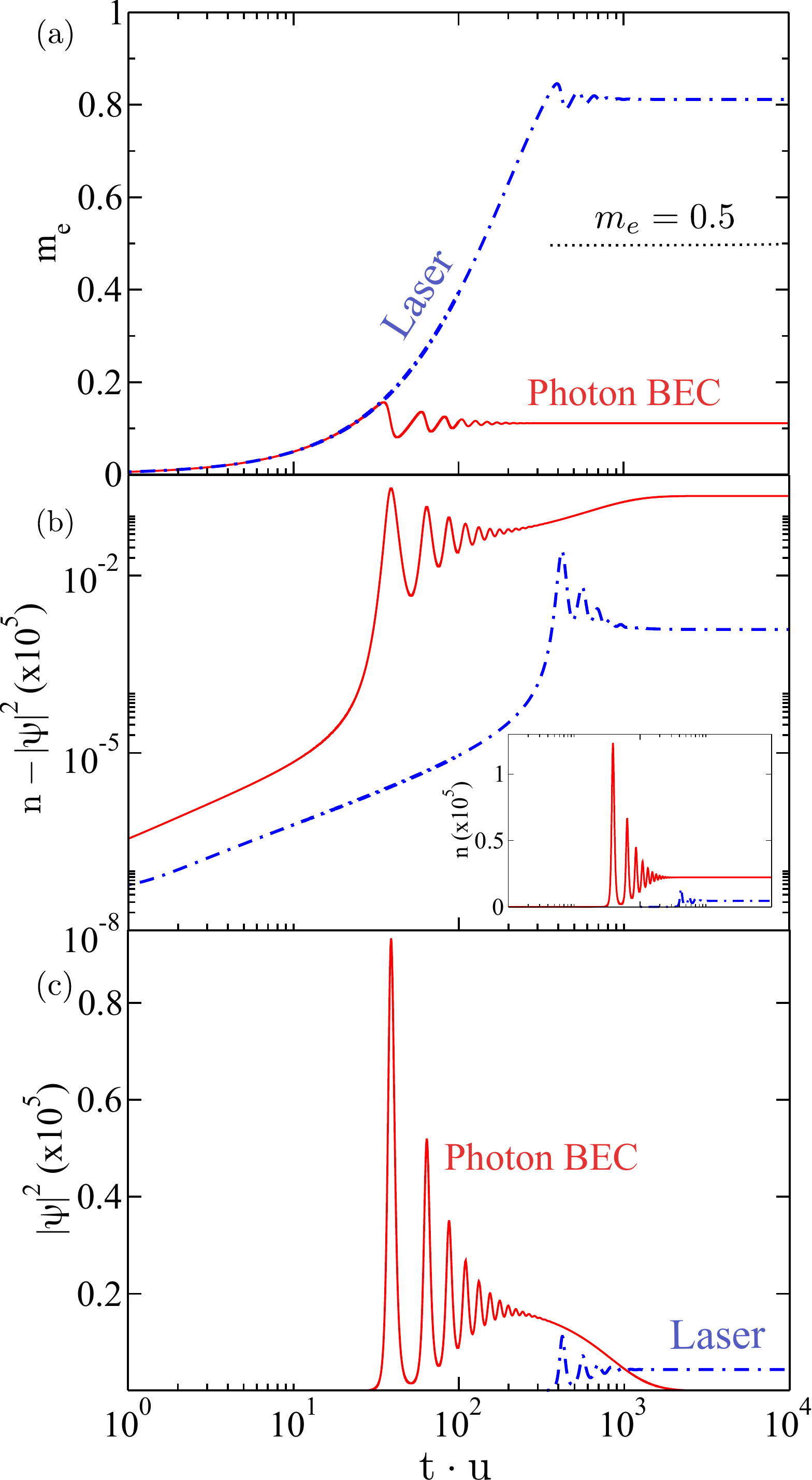}
\caption{{\it Time evolution.} Relaxation dynamics of the fraction of excited molecule $m_{e}$, the total photon number $n$, and the photon condensate density $|\psi|^2$ are shown in (a-c), respectively, for different ratios of absorption and emission, $\log_{10}\left(\Gamma_{\rm a}/\Gamma_{\rm e}\right)$=$-3.25$ (red, solid line) and $-2.38$ (blue, dash-dotted line), representing the condensate regime and a laser, respectively. We have set $M=10^7$, and the other parameters are the same as in Fig.~\ref{fig:time_evol}.} 
\label{fig:time_evol-GC}
\end{figure}

\sect{Grand-canonical versus canonical fluctuations}  
The results shown in the main text refer to a parameter regime of canonical statistical fluctuations, where the condensate population number $|\psi|^2$ in the plateau region is large compared to the photon fluctuation number $(n-|\psi|^2)$ in the metastable plateau region, see Fig.~\ref{fig:time_evol}.
However, the system can be tuned to a fluctuation-dominated regime where $(n-|\psi|^2)\approx |\psi|^2$, e.g., by reducing the molecule number $M$ (more dilute dye solution) or the (negative) cavity detuning, $\log_{10}(\Gamma_a/\Gamma_e)$ \cite{Schmitt14}. Fig.~\ref{fig:time_evol-GC} shows that for reduced molecule number $M=10^7$, keeping all other parameters as in Fig.~\ref{fig:time_evol}~\Circled{1}, the condensate and fluctuation oscillations during the initial BEC build-up period are enhanced, and the metastable BEC plateau is shorter and less pronounced. The oscillatory behavior is due to the presence of a non-Hermitian phase transition which has been found and analyzed in Ref.~\cite{Ozturk21}. In addition, the fluctuation number $(n-|\psi|^2)$ is of the same order of magnitude as $|\psi|^2$, and, due to the nonlinear dynamics, the number of molecule excitations $m_eM$ is much larger than $|\psi|^2$. This means that the photon gas is small compared to the reservoir of molecule excitations, that is, the system is in the fluctuation-dominated, grand-canonical regime.            

\begin{figure}[t]
\includegraphics[width=\columnwidth]{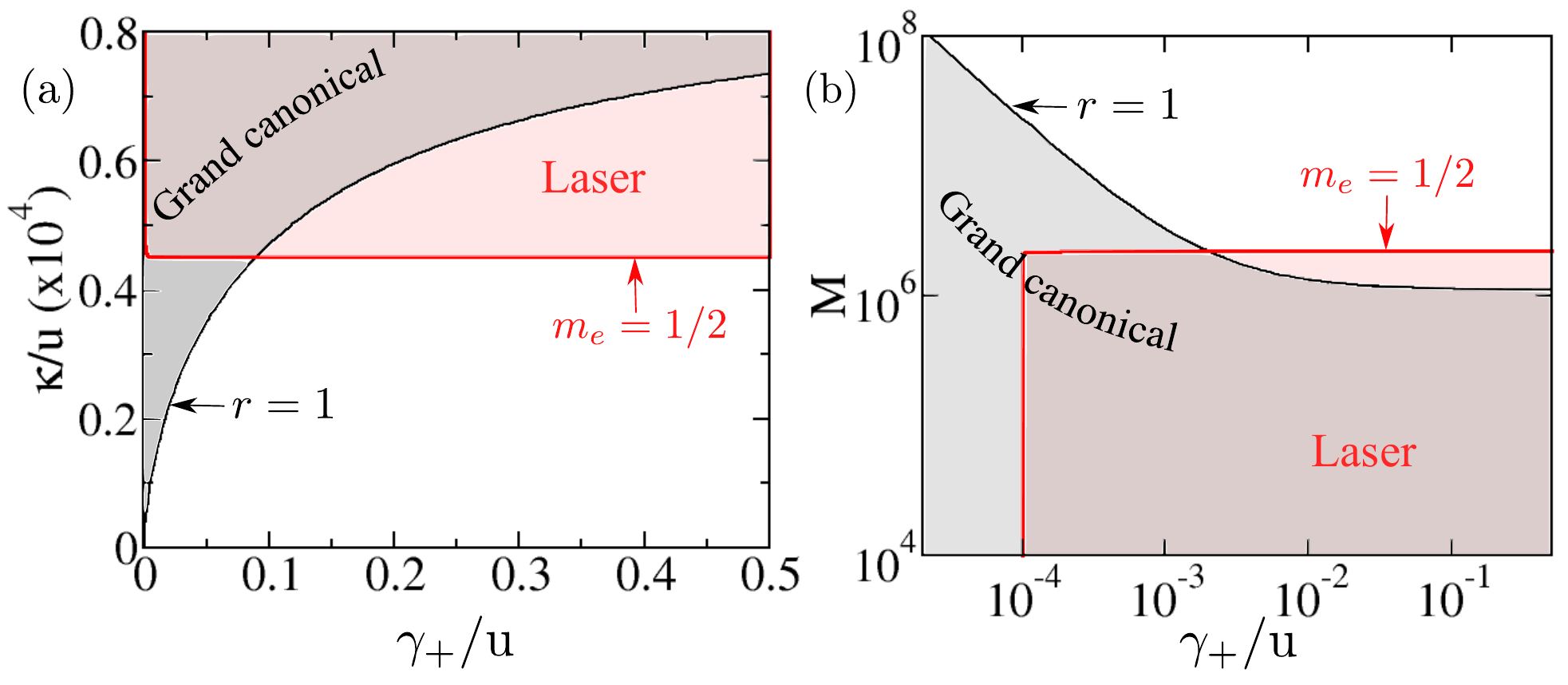}
\caption{{\it Tuning the grand canonical fluctuation.} The grand canonical regime with effective reservoir size $r_s>1$ are shown by gray-shaded areas in the (a) $\gamma_{+}/u$ vs $\kappa/u$ and (b) $\gamma_{+}/u$ vs $M$ plane. The laser regions are overlayed in the same plots as red-shaded areas enclosed by the transition line $m_e=1/2$. We have set $\log_{10}(\Gamma_{\rm a}/\Gamma_{\rm e})=-3.25$ (see \Circled{1} in Fig.~\ref{fig:phase_diagram}), and the other parameters are the same as in Fig.~\ref{fig:phase_diagram}.}
\label{fig:GC-boundary}
\end{figure}

In Fig.~\ref{fig:GC-boundary} we map out the canonical and grand-canonical regimes as defined by the relative reservoir size $r=M\overline{m}_e/\overline{n}^2$ \cite{Schmitt23}, its dependence on the external pump rate $\gamma_+$ and the cavity loss $\kappa$ or the dye-moleculae number $M$, respectively. $\overline{m}_e$ and $\overline{n}$ are calculated from the respective, stable FPs $\overline{X}_{\rm FP}^{0/\rm L}$. The canonical to grand-canonical crossover line is given by $r=1$. 

\begin{figure}[t]
\includegraphics[width=\columnwidth]{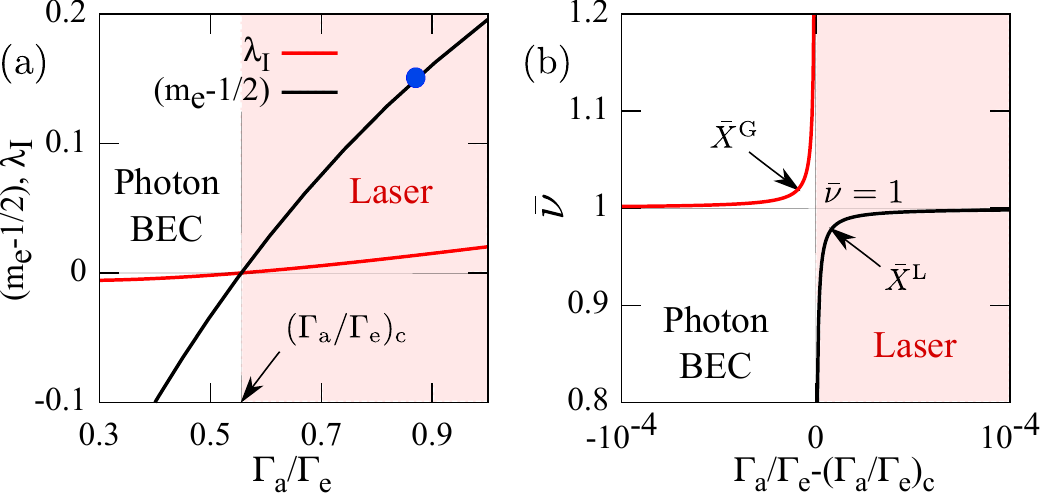}
\caption{{\it Transition to laser.} (a) The fraction of excited molecules $m_e$ computed at $t\cdot u=10^3$ from time evolution of the rate equations, and the instability exponent $\lambda_{\rm I}$ of the FP $\overline{X}^{0}$ are plotted as a function of $\Gamma_{\rm a}/\Gamma_{\rm e}$. (b) The condensate fraction $\overline{\nu}$ corresponding to $\overline{X}^{\rm L}$ and $\overline{X}^{\rm G}$ are plotted against $\Gamma_{\rm a}/\Gamma_{\rm e}-(\Gamma_{\rm a}/\Gamma_{\rm e})_c$, where the transition to laser occurs at $\Gamma_{\rm a}/\Gamma_{\rm e}=(\Gamma_{\rm a}/\Gamma_{\rm e})_c$. The blue dot marks the parameter values for the laser dynamics shown in Fig.\ref{fig:time_evol}. We set $\gamma_{+}/u=0.3$, and the other parameters are the same as in Fig.~\ref{fig:time_evol}.}
\label{fig:lasing-transition}
\end{figure}

\sect{Transition to laser}
In this section, we give details of the laser transition, as computed from the equations of motion and discussed in the main text. As shown in Fig.~\hyperref[fig:lasing-transition]{\ref{fig:lasing-transition}(a)}, the fraction of excited molecules, computed from numerical time evolution, undergoes a continuous change across the lasing threshold $m_e-1/2=0$ as a function of cavity detuning $\Gamma_a/\Gamma_e$. By contrast, the fixed points exhibit discontinuous behavior. The FP $\overline{X}^{0}$ becomes unstable at the transition to the laser, as depicted by the instability exponent $\lambda_{\rm I}={\rm max [Re(\lambda)]}$ computed from linear stability analysis around $\overline{X}^{0}$.

As the laser transition is approached from the photon BEC side, the ghost FP $\overline{X}^{\rm G}$ shifts more into the unphysical region, with $\overline{\nu}\to +\infty$ at the critical detuning $(\Gamma_a/\Gamma_e)_c$, see Fig.~\hyperref[fig:lasing-transition]{\ref{fig:lasing-transition}(b)}. Consequently, its influence on the physical dynamics vanishes. Instead, on the lasing side of the transition, a new, stable fixed point $\overline{X}^{\rm L}$ appears which is characterized by a physical condensate fraction $\overline{\nu} <1$ which continuously increases from $\overline{\nu}=0$ at the transition to $\overline{\nu}\to 1$ deep inside the lasing phase. Thus, the photon BEC and the laser are separated by a true nonequilibrium phase transition, as they are characterized by two distinct, stable fixed points. 

\begin{figure}[b]
\includegraphics[width=\columnwidth]{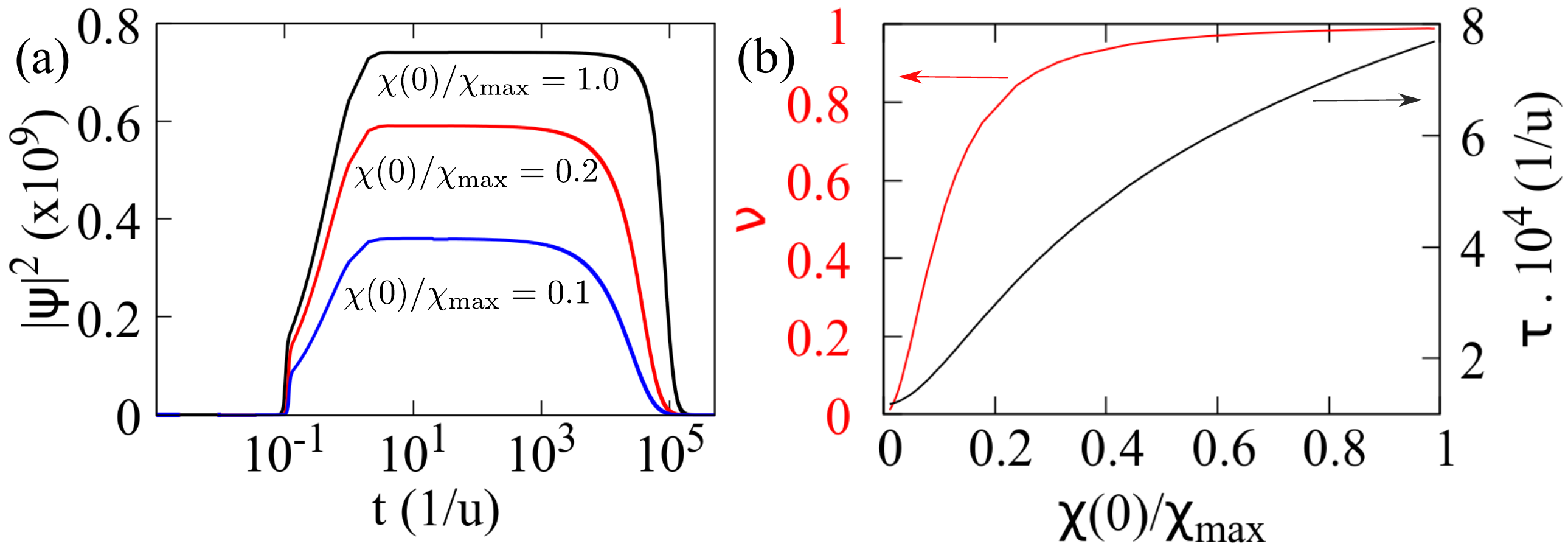}
\caption{{\it Dependence on radiative molecule emission amplitude.} (a) Time evolution of the condensate density $|\psi|^2$ for different initial value $\chi(0)/\chi_{\rm max}$. (b) The condensate lifetime $\tau$ and the condensate fraction $\nu$ are plotted as a function of $\chi(0)$. We have set $\log_{10}\left(\Gamma_{\rm a}/\Gamma_{\rm e}\right)=-3.25$, while the other parameters are the same as in Fig.~\ref{fig:time_evol}.}
\label{fig:condensate-vs-chi}
\end{figure}

\sect{Dependence on photon emission amplitude} Finally, we discuss the role of the radiative molecular transition amplitude $\chi$ in the dynamical stabilization of the condensate. 
The numerical solutions show a weak time dependence of $\chi$, so that it can be characterized by its initial value. 
In Fig.~\hyperref[fig:condensate-vs-chi]{\ref*{fig:condensate-vs-chi}(a)}, we plot the condensate evolution for different initial values of $\chi(0)/\chi_{\rm max}$, c.f. main text. For the maximal value $\chi(0)/\chi_{\rm max}=1$, we find up to $>95\%$ photons in the condensate, which also corresponds to the experimental observation \cite{Weitz10}. With decreasing $\chi(0)/\chi_{\rm max}$, we observe a decrease in the condensate fraction. We also determined the condensate lifetime $\tau$, measured as the time when the condensate density becomes half of its plateau value, see Fig.~\hyperref[fig:condensate-vs-chi]{\ref*{fig:condensate-vs-chi}(b)}. In case of a ghost-attractor, the bottleneck time-scale can, more formally, be computed from the distance of the trajectory to the ghost-attractor \cite{Strogatz-book}. As can be noticed, for $\chi(0)/\chi_{\rm max}=1$, the half-life becomes $\tau\sim 10^5$, which corresponds to the lowest positive eigenvalue obtained by performing linear stability analysis around the FP $\overline{X}_{\rm FP}^{\rm G}$.

\vfill
\eject

\clearpage
\onecolumngrid
\pagestyle{empty}

\renewcommand{\thefigure}{S\arabic{figure}}
\setcounter{figure}{0}
\renewcommand{\theequation}{S-\arabic{equation}}
\setcounter{equation}{0}

\renewcommand{\thefootnote}{\fnsymbol{footnote}}

\Cn{{\large \bf Supplemental Material}}
\vspace{-0.2cm}

\maketitle

\onecolumngrid

\vspace{-1cm}
\Cn{$^a$These authors contributed equally to this work.}

In this Supplement we present details on the derivation of the fixed points (FPs) and their linear stability discussed in the main text and in the End Matter, as well as the photon BEC dynamics for low photon number. 

\section{Fixed Points}
\label{Supp-FP}

The steady-states of the system in the long time can be obtained by setting the time derivatives of the rate equations, see Eq.~\eqref{eq:rate_eqn} in the main text, to zero, i.e.  $\partial_t \psi = 0 = \partial_t \chi$ and $\partial_t m_e = 0 = \partial_t n$, which yields,
\begin{subequations}
    \begin{eqnarray}
        \partial_t{\psi} &=& 0 = -ig_{\beta} M \chi + \left[MG_{\psi}(m_e)-\kappa\right]\psi/2 
        \label{SS:psi}
        \\
        \partial_t{\psi} &=& 0 = i g_{\beta} \psi (2m_e-1) - G_{\chi}(n)\chi/2
        \label{SS:chi}
        \\
        \partial_t{m_e} &=& 0 = - 2 g_{\beta} {\rm Im}[\psi^* \chi] + \gamma_{+}(1-m_e) - \gamma_{-} m_e - R(m_e,n)
        \label{SS:me}
        \\
        \partial_t{n} &=& 0 = 2 g_{\beta} M {\rm Im}[\psi^* \chi] - \kappa n + MR(m_e,n)
        \label{SS:n}
    \end{eqnarray}
    \label{SS}%
\end{subequations}
In order to obtain the steady-state solution $\overline{X}=(\overline{\psi},\overline{\chi},\overline{m}_e,\overline{n})$ from Eq.~\eqref{SS}, we first express $\overline{\chi}$ in terms of $\overline{\psi}$ from Eq.~\eqref{SS:psi}, which gives the following relation,
\begin{equation}
    \overline{\chi} = \frac{MG_{\psi}(\overline{m}_e)-\kappa}{2ig_{\beta}M}~\overline{\psi}
    \label{chi-psi}
\end{equation}
Inserting the expression of $\overline{\chi}$ in Eq.~\eqref{SS:n}, and using the relation $R(\overline{m}_e,\overline{n})=\overline{n}G_{\psi}(\overline{m}_e)+\Gamma_{\rm e}\overline{m}_e$, we obtain the condensate fraction,
\begin{equation}
\overline{\nu} = \frac{|\overline{\psi}|^2}{\overline{n}} = 
1-\frac{M\Gamma_e \overline{m}_e}{\overline{n}[\kappa-MG_{\psi(\overline{m}_e}]}\ , 
\label{Seq:nu}
\end{equation}
as written in the main text, see Eq.~\eqref{eq:nu}. Also, multiplying Eq.~\eqref{SS:me} by $M$ and adding that to Eq.~\eqref{SS:n}, we arrive at a general relation,
\begin{equation}
    \overline{m}_e = \left( \gamma_+ - \frac{\kappa \overline{n}}{M} \right)/(\gamma_+ + \gamma_-),
\label{Seq:m_e-n}
\end{equation}
see Eq.~\eqref{eq:m_e-n} in the main text. It can be noticed, that Eq.~\eqref{SS:psi} and Eq.~\eqref{SS:chi} admit a vanishing condensate solution, i.e. $\overline{\psi}=0=\overline{\chi}$. Putting it back in Eqs.~\eqref{SS:me} and \eqref{SS:n}, and solving for $\overline{n}$ gives the solution,
\begin{equation}
    \overline{n}^0 = -\frac{A}{2B} + \sqrt{\left(\frac{A}{2B}\right)^2+\frac{\gamma_{+}\Gamma_{\rm e} M}{B}} 
\label{Seq:nnull}
\end{equation}
where, $A = \left(\gamma_{-} \Gamma_{\rm a} - \gamma_{+} \Gamma_{\rm e}\right) M +\kappa \left(\gamma_{+}+\gamma_{-}+\Gamma_{\rm e}\right)$ and $B=\kappa (\Gamma_{\rm a}+\Gamma_{\rm e})$. The corresponding fraction of excited molecules $\overline{m}_e=\overline{m}_e^0$ can be obtained from Eq.~\eqref{Seq:m_e-n}. Thus, we obtain the fixed point with vanishing condensate as $\overline{X}=\overline{X}^0=(0,0,\overline{m}_e^0,\overline{n}^0)$.

The nonvanishing solution of $\overline{\psi}$ and $\overline{\chi}$ can be obtained by solving Eqs.~\eqref{SS:psi} and \eqref{SS:chi} as,
\begin{equation}
    \begin{bmatrix}
        \left(MG_{\psi}(\overline{m}_e)-\kappa\right)/2 & -ig_{\beta}M \\
        ig_{\beta}(2\overline{m}_e-1) & -G_{\chi}(\overline{n})/2
    \end{bmatrix}
    \begin{bmatrix}
        \overline{\psi} \\
        \overline{\chi}
    \end{bmatrix}
    = 0
    \label{Eq:psi-chi}
\end{equation}
For a solution $\overline{\psi}\neq 0 \neq \overline{\chi}$, one requires the determinant of Eq.~\eqref{Eq:psi-chi} to be zero, which yields,
\begin{equation}
    4 g_{\beta}^2 M (2\overline{m}_e-1) - G_{\chi}(\overline{n}) \left[\kappa-MG_{\psi}(\overline{m}_e)\right] = 0,
\label{Seq:determinant} 
\end{equation}
see Eq.~\eqref{eq:determinant} in the main text. By inserting the expression of $\overline{m}_e$ from Eq.~\eqref{Seq:m_e-n} into Eq.~\eqref{Seq:determinant}, one obtains the solutions, $\overline{n}=\overline{n}^{\rm G}$ and, by Eq.~\eqref{Seq:m_e-n}, $\overline{m}_e=\overline{m}_e^{\rm G}$. In the weakly pumped case, which is our regime of interest, we obtain $2\overline{m}_e-1<0$, indicating no population inversion. Thus, it follows from Eq.~\eqref{Seq:determinant}, that $\left[\kappa-MG_{\psi}(\overline{m}_e^{\rm G})\right]<0$ since $G_{\chi}(\overline{n})>0$. From the expression of $\overline{\nu}$ in Eq.~\eqref{Seq:nu}, it, thus, follows $\overline{\nu}=\overline{\nu}^{\rm G}>1$. We refer this fixed point $\overline{X}^{\rm G}$ as ghost-attractor in the main text. As shown in Fig.~A3, that the ghost FP moves away from the physical realm as the laser transition is approached. At the laser transition, a new solution of Eq.~\eqref{Seq:determinant}, $\overline{n}=\overline{n}^{\rm L}$ and $\overline{m}_e=\overline{m}_e^{\rm L}$, appears with population inversion in molecules, i.e. $2\overline{m}_e^{\rm L}-1>0$, as well as $\overline{\psi}^{\rm L}\neq 0\neq \overline{\chi}^{\rm L}$, leading to a physical solution $\overline{\nu}^{\rm L}<1$ of Eq.~\eqref{Seq:nu}. The corresponding FP $\overline{X}^{\rm L}=\{\overline{\psi}^{\rm L},\overline{\chi}^{\rm L},\overline{m}_e^{\rm L},\overline{n}^{\rm L}\}$ is referred to the lasing state in the main text.

Since $\overline{\psi}$ and $\overline{\chi}$ are complex numbers, we parametrize them by their amplitudes and phases as $|\overline{\psi}|e^{i\overline{\theta}_{\psi}}$ and $|\overline{\chi}|e^{i\overline{\theta}_{\chi}}$. Using the solutions of $\overline{n}$ and $\overline{m}_e$ we obtain $|\overline{\psi}|$ from Eq.~\eqref{Seq:nu}, and $|\overline{\chi}|$ from Eq.~\eqref{chi-psi}. Since the equations of $\partial_t\psi$ and $\partial_t\chi$ are invariant under a gauge transformation, the global phase $(\overline{\theta}_{\psi}+\overline{\theta}_{\chi})$ remains undetermined, however, the relative phase $(\overline{\theta}_{\psi}-\overline{\theta}_{\chi})$ can be obtained from $\partial_t n=0$ by using the values of other variables in the steady state. A few steps of algebra yields a phase relation,
\begin{equation}
\sin\left(\overline{\theta}_{\psi}-\overline{\theta}_{\chi}\right) = 1 \Rightarrow  \left(\overline{\theta}_{\psi}-\overline{\theta}_{\chi}\right)=\pi/2.    
\end{equation}
We would like to emphasize, that the phase relation is not only true in the steady states, but also is followed in the dynamically stable condensate regime. 

\section{Stability Analysis}
\label{Supp-Stability}

We analyze the linear stability of the fixed points discussed in the previous section. It is performed by considering the deviations from the FPs, $X(t)=\overline{X}+\Delta X\, e^{\lambda t}$, which, after inserting into Eq.~\eqref{eq:rate_eqn} (in the main text) and expanding up to the linear order, leads to an eigenvalue equation, $(\mathcal{M}-\lambda\, \mathds{1})\,\Delta X=0$ with the stability matrix $\mathcal{M}$, the Lyapunov exponent $\lambda$, and $\Delta X=(\Delta \psi,\Delta\psi^*,\Delta\chi,\Delta\chi^*,\Delta m_e,\Delta n)^{\rm T}$. The linearized stability matrix $\mathcal{M}$ is given by,
\begin{equation}
    \mathcal{M}=\begin{pmatrix}
        -\frac{[\kappa - MG_{\psi}(\overline{m}_e)]}{2} & 0 & -i g_\beta M & 0 & \frac{\Gamma_{\rm a} +\Gamma_{\rm e}}{2} M \overline{\psi} & 0 \\
        0 & -\frac{[\kappa - MG_{\psi}(\overline{m}_e)]}{2} & 0 & i g_\beta M & \frac{\Gamma_{\rm a} +\Gamma_{\rm e}}{2} M \overline{\psi}^* & 0 \\
        i g_\beta (2 \overline{m}_e-1) & 0 & -\frac{G_{\chi}(\overline{n})}{2} & 0 & i 2 g_\beta \overline{\psi} & -\frac{\Gamma_{\rm a}+\Gamma_{\rm e}}{2} \overline{\chi} \\
        0 & -i g_\beta (2 \overline{m}_e-1) & 0 & -\frac{G_{\chi}(\overline{n})}{2} & -i 2 g_\beta \overline{\psi}^* & -\frac{\Gamma_{\rm a}+\Gamma_{\rm e}}{2} \overline{\chi}^* \\
        -i g_\beta \overline{\chi}^* & i g_\beta \overline{\chi} & i g_\beta \overline{\psi}^* & - i g_\beta \overline{\psi} & -G_{\chi}(\overline{n}) & -G_{\psi}(\overline{m}_e) \\
        i g_\beta M \overline{\chi}^* & -i g_\beta M\overline{\chi} & -i g_\beta M\overline{\psi}^* &  i g_\beta M \overline{\psi} & M\left[\Gamma_{\rm a}\overline{n}+\Gamma_{\rm e}(\overline{n}+1) \right] & -[\kappa - MG_{\psi}(\overline{m}_e)]
    \end{pmatrix}.
\end{equation}
We diagonalize $\mathcal{M}$ and obtain the eigenvalues $\lambda$. Stability of a FP is determined if all the eigenvalues have ${\rm Re}(\lambda)<0$. For the ghost FP $\overline{X}^{\rm G}$ discussed in Fig.~\ref{fig:flow-diagram}, we find a single positive Lyapunov exponent ${\rm Re}(\lambda)={\rm Re}(\lambda^{\rm G}) \approx 5.89\times 10^{-5}u$, and the other eigenvalues have negative real part ${\rm Re}(\lambda)<0$ with the smallest one in magnitude being ${\rm Re}(\lambda)\approx -2u$. The latter one has the normalized eigenvector with dominant contribution from $\Delta n \approx 1$, whereas, the eigenvector corresponding to $\lambda^{\rm G}$ has the dominant contribution from $\Delta \psi = \Delta \psi^* \approx 1/\sqrt{2}$. These two eigenvectors are plotted in the $n-\nu$ plane in Fig.~\ref{fig:flow-diagram} in the main text. The plateau time scale for the photon BEC, computed numerically agrees well with the Lyapunov exponent ${\rm Re}(\lambda^{\rm G})$, see the discussions in the main text and End Matter.

\section{Time evolution for lower photon number}

\begin{figure}[ht]
\includegraphics[width=\columnwidth]{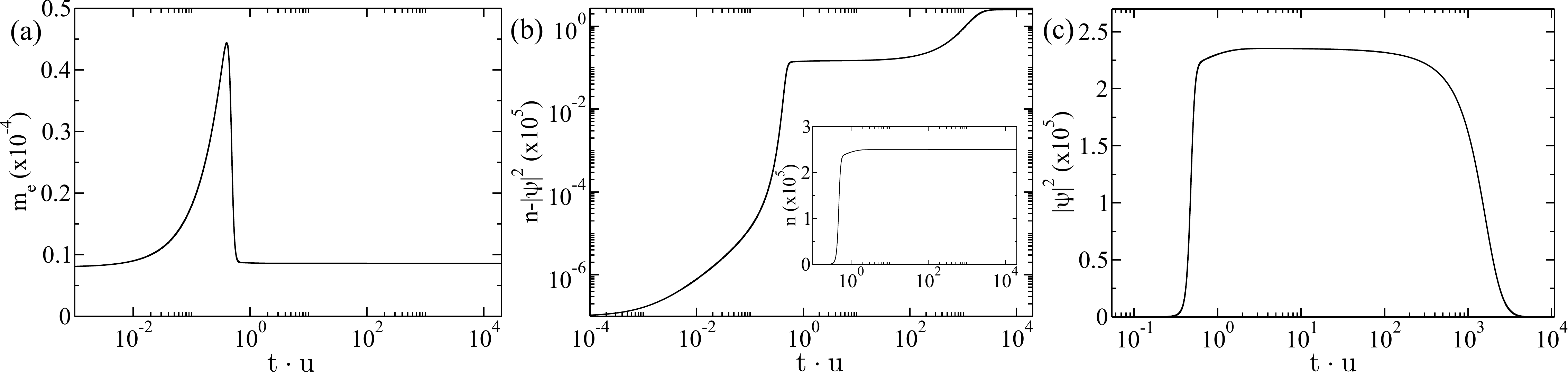}
\caption{{\it Time evolution.} Relaxation dynamics of (a) the fraction of excited molecule $m_{e}$, (b) non-condensed photon number ($n-|\psi|^2$) with total photon number $n$ in the inset, and (c) photon condensate density $|\psi|^2$ are shown for lower pumping rate $\gamma_+/u=10^{-4}$ and ratios of absorption and emission, $\log_{10}\left(\Gamma_{\rm a}/\Gamma_{\rm e}\right)$=$-5.21$. Note the condensate fraction at the plateau is $\sim 95\%$ in this case. We set the initial $m_{\rm e}(0)=0.0008\%$, while the other parameters and initial conditions for the time evolution are the same as in Fig.~\ref{fig:time_evol}.} 
\label{fig:time_evol-low-pump}
\end{figure}

Here we present the time evolution of photon number $n$, condensate density $|\psi|^2$, non-condensed photons $n-|\psi|^2$, and the excited molecule fraction $m_{\rm e}$ at a lower pumping relevant for existing photon BEC experiments using Rhodamine 6G dye molecules, see Fig.~\ref{fig:time_evol-low-pump}. It can be noted, that the time evolution with lower photon number remains qualitatively same to that presented in Fig.~\ref{fig:time_evol}.

\end{document}